\documentclass[useAMS,usenatbib]{mn2e}
\usepackage{natbib}
\usepackage{graphicx}

\title[Numerical modeling of Palomar 4]{Direct $N$-body simulations of globular
clusters - II.  Palomar 4}

\author[Zonoozi et al.]
{Akram Hasani Zonoozi$^{1}$\thanks{
E-mail:  \mbox{a.hasani@iasbs.ac.ir} (AHZ)
\mbox{haghi@iasbs.ac.ir} (HH);
\mbox{akuepper@astro.columbia.edu}(AHWK);
 \mbox{h.baumgardt@uq.edu.au}(HB);
  \mbox{mfrank@lsw.uni-heidelberg.de } (MF);
 \mbox{pavel@astro.uni-bonn.de} (PK)
 }
, Hosein Haghi$^{1}$, Andreas H.W. K\"{u}pper$^{2}$\thanks{Hubble Fellow},
\newauthor
Holger Baumgardt$^{3}$, Matthias J. Frank$^4$, Pavel Kroupa$^5$\\
$^{1}$Department of Physics, Institute for Advanced Studies in Basic Sciences (IASBS), PO Box 11365-9161, Zanjan, Iran\\
$^{2}$Department of Astronomy, Columbia University, 550 West 120th Street, New York, NY 10027, USA\\
$^{3}$School of Mathematics and Physics, University of Queensland, Brisbane, QLD 4072, Australia,\\
$^{4}$Landessternwarte, Zentrum f\"ur Astronomie der Universit\"at Heidelberg, K\"onigsstuhl 12, D-69117 Heidelberg, Germany\\
$^{5}$Helmhotz-Institit f\"ur Strahlen-und Kernphysik (HISKP), Universit\"at Bonn, Nussallee 14-16, D-53115 Bonn, Germany\\}

\begin{document}

\date{Accepted 2014 March 13. Received 2014 March 13; in original form 2014 January 26}

\pagerange{\pageref{firstpage}--\pageref{lastpage}} \pubyear{2013}

\maketitle

\label{firstpage}

\maketitle

\begin{abstract}
We use direct $N$-body calculations to study the evolution of the unusually extended outer halo globular cluster Palomar 4 (Pal~4) over its entire lifetime in order to reproduce its observed mass, half-light radius, velocity dispersion and mass function slope at different radii.

We find that models evolving on circular orbits, and starting from a non-mass segregated, canonical initial mass function (IMF) can reproduce neither Pal 4's overall mass function slope nor the observed amount of mass segregation. Including either primordial mass segregation or initially flattened IMFs does not reproduce the observed amount of mass segregation and mass function flattening simultaneously. Unresolved binaries cannot reconcile this discrepancy either. We find that only models with both a flattened IMF \textit{and} primordial segregation are able to fit the observations. The initial (i.e. after gas expulsion) mass and half-mass radius of Pal~4 in this case are about 57000 M${\odot}$ and 10 pc, respectively. This configuration is more extended than most globular clusters we observe, showing that the conditions under which Pal~4 formed must have been significantly different from that of the majority of globular clusters. We discuss possible scenarios for such an unusual configuration of Pal~4 in its early years.

\end{abstract}


\section{Introduction}

Globular clusters are ideal astrophysical systems whose long-term
evolution is determined by several internal and external processes,
like mass loss due to stellar evolution and the energy-equipartition processes as well as tidal removal of star. In this regard, numerous
numerical investigations have been carried out to understand their dynamical evolution
(e.g.~\citealt{Giersz11}; see also the textbook by Heggie \& Hut 2003). However, only within the last few
years,  with the introduction of graphics processing unit (GPU)-accelerated $N$-body codes
such as \textsc{nbody6} \citep{Aarseth03, Nitadori12} it has
become feasible to compute the dynamical evolution of massive
star clusters over their entire lifetimes on a star-by-star
basis. In Paper I of this series \citep{Zonoozi11}, we presented
the first direct $N$-body simulation of a Milky Way globular
cluster over a Hubble time. For this project we chose the
outer-halo globular cluster Palomar~14 (Pal~14), due to its relatively low
mass and its large half-mass radius. In the paper, we presented a comprehensive  set of
$N$-body computations of Pal~14's evolution over its entire lifetime
and compared the results to the observed mass, half-light radius, flattened stellar mass function and
velocity dispersion of Pal~14, which have been presented by Jordi et al.
(2009). We showed that dynamical mass segregation alone cannot explain the mass function flattening in the cluster centre when
starting from a canonical Kroupa initial mass function (IMF), and that a very high degree
of primordial mass segregation would be necessary to explain this
discrepancy. We concluded that such initial conditions for Pal~14
might be obtained by a violent early gas-expulsion phase from an
embedded cluster born with mass segregation and a canonical IMF for low-mass stars, a thesis supported later by an independent study of an ensemble of globular clusters \citep{Marks08}.

Here we aim at modelling the globular cluster Palomar~4 (Pal~4), which is similar to Pal~14
but has more complete observational data.
Recently, Frank et al. (2012) presented an extensive observational study of Pal~4, revealing a flattened stellar mass function and significant mass segregation throughout the cluster. This additional knowledge of Pal~4 puts much stronger constraints on its current dynamical state.

Star clusters can undergo significant changes not only at birth but also during
the course of their dynamical evolutions. It is therefore essential to
specify to what extent the present-day properties of a globular
cluster, e.g.~their degree of mass segregation, are imprinted by early evolution and the formation
processes, and to what extent they are the outcome of long-term
dynamical evolution.

There are certain distinct mechanisms that can cause mass segregation.
Dynamical mass segregation is the process by which the  more
massive stars of a gravitationally bound system sink towards the
central regions, while lighter stars move further away from the
centre. This process is a consequence of evolution towards energy equipartition
driven by two-body encounters and is usually associated with the
long-term evolution of clusters through the two-body relaxation process.

However, a number of observational studies  (e.g.~\citealt{Hillenbrand97, Fischer98, Hillenbrand98, de Grijs02, Sirianni02, Gouliermis04, Stolte06, Sabbi08, Allison09, Gouliermis09}) have found evidence of mass segregation in clusters with ages shorter than the time needed to produce the observed
segregation via two-body relaxation (see also~\citealt{de Grijs10}).

It has been suggested that
the observed mass segregation in young clusters could be
primordial -- imprinted by the early star-formation process
\citep{Bonnell97, Bonnell01, Bonnell98, Klessen01, Bonnell06}. Such mass segregation could be due to the
higher accretion rate of proto-stars in high-density regions of
molecular clouds fragmenting into clumps.  If individual clumps
are mass segregated, it has been shown by McMillan, Vesperini,
\& Portegies Zwart (2007), that such primordial mass segregation would not be erased in the
violent-relaxation phase during which clumps merge. The final
system would preserve the mass segregation of the original clumps
(see also \citealt{Fellhauer09, Moeckel09}).
But even if such clumps are not initially segregated, the internal
segregation time-scale can be shorter than the time needed for
the clumps to merge. Hence, they will segregate internally via
two-body relaxation and preserve this segregation after they have
merged (\citealt{McMillan07}).

Alternatively, Bastian et al. (2008) found observational evidence for a
strong expansion in the first 20 Myr of the evolution of six young M51
clusters and pointed out that this expansion could also lead to a
rapid variation in the cluster relaxation time, thus, using the present-day relaxation time might lead to
an underestimation of the possible role played by two-body
relaxation in generating mass segregation in the early phases of
a cluster's dynamical evolution.

Regardless of the mechanism producing mass segregation, the
presence of primordial (or early) mass segregation significantly
affects the global dynamical evolution of star clusters (\citealt{Gurkan04, Baumgardt08a}). For example, the early mass
loss due to stellar evolution of high-mass stars has a stronger
impact on initially segregated clusters than on unsegregated
clusters (\citealt{Vesperini09b}). The degree of primordial or
early mass segregation is therefore a crucial parameter in the
modelling of globular clusters.

Another important quantity that has to be taken into account in the
modelling of star clusters is the IMF. Its shape strongly
influences the dynamical evolution of star clusters. The canonical
IMF as observed in young star clusters in the Milky Way is often expressed as a two-part
power-law function ($\frac{dN}{dm}\propto m^{-\alpha}$) with near
Salpeter-like slope above $0.5\,\mbox{M}{\odot}$ (i.e.,
$\alpha=2.3$; \citealt{Salpeter55}), and a shallower slope  of $\alpha=1.3$ for stars
in the mass range $0.08-0.5\,\mbox{M}{\odot}$
\citep{Kroupa01, Kroupa08, Kroupa13}.

The mass function of stars in clusters
evolves through stellar evolution and through dynamical evolution, i.e. via preferential
loss of low-mass stars \citep{Vesperini97,Baumgardt03}. This effect
should be more pronounced in concentrated clusters, since the
two-body relaxation times-cale is shorter for such systems.
However, based on a data set of observed mass functions of a
sample of globular clusters, De Marchi et al. (2007) found that
all high concentration clusters have steep mass functions (i.e.,
larger $\alpha$), while low concentration ones have a smaller
$\alpha$, although the opposite is expected.  This
effect is not well understood yet. Marks et al. (2008) suggested
that the `De Marchi relation' is due to early gas
expulsion. They showed that for initially mass-segregated
clusters mostly low-mass stars are lost due to gas expulsion,
which yields a shallower slope in the low-mass range in clusters with low concentration.

Moreover, the mass functions of some outer-halo globular clusters
also show a flattening at comparatively high stellar masses (i.e.,
the range $0.55\leq m/M{\odot}\leq0.85$; \citealt{Jordi09, Frank12}).
Some of the ideas that have been proposed to
explain a shallowness of the slope at the high-mass end, again,
include primordial mass segregation of stars in the cluster (e.g.,
\citealt{Vesperini97, Kroupa02, Mouri02}).
It remains to be shown if such scenarios can really reproduce
the observational findings. Direct $N$-body simulations offer the possibility to test these scenarios.

In this paper we perform a set of
direct $N$-body simulations of Pal~4 to determine its most
likely initial conditions in terms of total mass, initial
half-mass radius, stellar mass function and primordial mass
segregation. We furthermore investigate the effect of unresolved
binaries on the observed mass function of this cluster.

The paper is organized as follows. In
Section~\ref{Sec:Observational data} we describe the
observational data of Pal~4, including the velocity dispersion,
the mass function and the total stellar mass to which we later compare our simulations. In Section~\ref{Sec:Description of
the models} we describe the set-up of the $N$-body models. This
is followed by a presentation of the results of simulations for
different scenarios in Section~\ref{Sec:Results}. A discussion and
conclusions are presented in Section~\ref{Sec:Conclusions}.

\section{Observational data}\label{Sec:Observational data}

We compare the results of our numerical modelling of Pal~4 with
the observational data by \citet{Frank12}, who have presented
a spectroscopic and photometric study of Pal~4.

\citet{Frank12} have determined Pal~4's velocity dispersion by
measuring the line-of-sight radial velocities of a sample of
member stars using the High Resolution Echelle Spectrograph
(HIRES; \citealt{Vogt94}) mounted on the Keck I telescope. Using
the radial velocities of 23 clean member stars (excluding an outlier called \emph{star 12}) the internal radial velocity dispersion of Pal~4 was measured to be $\sigma=0.87\pm0.18$ kms$^{-1}$.

Frank et al.~also obtained the cluster's mass function down to a
limiting magnitude of $V\approx28$\,mag using \emph{Hubble Space
Telescope}/Wide Field Planetary Camera 2 (\emph{HST}/WFPC2) data from the
HST archive. They determined the stellar mass function of the
cluster in the mass range $0.55\leq m/M{\odot}\leq0.85$,
corresponding to stars from the tip of the red giant branch (RGB) down to the 50 per
cent completeness limit in the cluster's core at the faint end,
removing likely foreground stars, blue stragglers and horizontal
branch stars. The best-fitting present-day mass function slope of Pal~4 was
found to be $\alpha=1.4\pm0.25$. This is significantly shallower
than a \citet{Kroupa01} IMF with $\alpha=2.3$ in this range of
masses, and is similar to the mass function in other Galactic globular clusters
(e.g \citealt{De Marchi07, Jordi09, Paust10}).
Moreover, Frank et al.~found that the slope of the mass function
steepens with radius from a slope of $\alpha \leq 1$ inside $r \leq 1.3r_h$ to $\alpha \geq 2.3$ at the largest observed radii, indicating the presence of mass segregation in Pal~4. Since the two-body relaxation time of Pal~4 is of the order of a Hubble time, the authors concluded that this may be an indication for primordial mass segregation.

Frank et al. (2012) estimated the total stellar mass inside the projected radius
covered by the WFPC2 pointing, $r<2.26$\,arcmin, and in the
stellar mass range  $0.55\leq m/\mbox{M}{\odot} \leq0.85$
to be $5960\pm110\,\mbox{M}{\odot}$, without considering the
contribution of blue stragglers and horizontal branch stars,
which is negligible due to their low number. Using the measured
slope for masses down to $0.5\, \mbox{M}{\odot}$ and adopting a
\citet{Kroupa01} mass function, with $\alpha=1.3$ for masses
between $0.08$ and $0.5\,\mbox{M}{\odot}$ and
$\alpha=0.3$ for masses between $0.01$ and
$0.08\,\mbox{M}{\odot}$, Frank et al.~estimated the mass of
Pal~4 to be about $M=14500\pm1300\,\mbox{M}{\odot}$ in the mass
range $0.01\leq m/\mbox{M}{\odot}\leq 0.85$.
Extrapolating out to the tidal
radius ($r_t=3.90\pm0.20$\,arcmin), and including the contribution of
remnant white dwarfs, they finally derive a total cluster mass of
$M_{tot}=29800\pm800\,\mbox{M}{\odot}$.

According to \citet{Frank12}, Pal~4's distance  from the Sun is
102.8 $\pm$ 2.4\,kpc. This is slightly closer than the 109.2\,kpc
given by Harris (1996, edition 2010), but well inside the range
of other previous estimate, by e.g. \citet{Burbidge58}, \cite{Christian86} and \citet{VandenBerg00} who derived
distances of 100, 105$\pm$5, and 104\,kpc, respectively. Assuming a circular velocity of 220\,km/s, the orbital period of Pal~4 around Galaxy is about 3 Gyr.

Frank et al. (2012) estimate an age of $11\pm 1$\,Gyr for
Pal~4 by adopting [Fe/H]$ = -1.41\pm 0.17$\,dex for the
metallicity from the best-fitting isochrone of \citet{Dotter08}.

Regarding the surface brightness profile of the cluster, the
best-fitting King (1966) model [based on the WFPC2 data and on
broad-band imaging with the Low-Resolution Imaging Spectrometer
(LRIS) at the Keck II telescope] yields a core
radius of $r_{c}=0.43\pm0.03 $\,arcmin, corresponding to $12.8\pm1.1$\,pc and a tidal radius of
$r_{t}=3.90\pm0.20$\,arcmin, corresponding to $115.5\pm10.2$\,pc. As a result, the concentration
parameter $c = \rm log_{10}(r_t/r_c)$ is $0.96\pm0.04$ and the projected
half-light radius is $r_{hl}=0.62\pm0.03$\,arcmin, corresponding to
$18.4\pm 1.1$\,pc. The corresponding 3D half-light radius is
about 24 pc.

In order to find the most likely initial conditions which
reproduce these observational values, e.g. projected half-light
radius, $r_{hl}$, radial velocity dispersion of stars,
$\sigma_{los}$, and mass function slope, $\alpha$, we construct a
set of $N$-body models for Pal~4 in the next section.

\section{Description of the models}\label{Sec:Description of the models}

In order to find the most likely initial conditions which
reproduce these observational values, e.g. projected half-light
radius, $R_{hl}$, radial velocity dispersion of stars,
$\sigma_{los}$, and mass function slope, $\alpha$, we construct a
set of $N$-body models for Pal~4. We use the collisional $N$-body code
\textsc{nbody6} \citep{Aarseth03, Nitadori12} on the GPU
computers of the University of Queensland to compute a
comprehensive set of initial models of Pal~4 over its entire
lifetime. \textsc{nbody6} uses a fourth-order Hermite integration
scheme, an individual time step algorithm to follow the orbits
of cluster members, invokes regularization schemes to deal
with the internal evolution of small-$N$ subsystems, and treats stellar evolution by including analytical fitting functions (see \citealt{Aarseth03} for details). It is the most advanced computer code available for our propose.

All clusters were set up using the publicly available code
\textsc{McLuster}\footnote{\tt https://github.com/ahwkuepper/mcluster}
\citep{Kupper11}. We simulate clusters consisting initially of
$N\simeq10^5$ stars with positions and velocities chosen
according to a Plummer profile \citep{Plummer11} in virial
equilibrium. The initial half-light radius and mass of the
cluster are chosen from an appropriate range, such that the
simulated clusters have projected half-light radii and masses at
11 Gyr close to the observed values for Pal~4. Since the
half-mass radius increases owing to mass loss by stellar evolution and dynamical evolution over the cluster lifetime, we chose
the initial 3D half-mass radii in the range of 8 to 15 pc to
reach 3D half-light radii of about 24 pc after 11 Gyr.

We first start with the canonical Kroupa IMF \citep{Kroupa01} to
assign masses to stars using lower and upper mass limits of
0.08 and 100\,M${\odot}$, respectively. For each
cluster profile we perform two sets of simulations, one with
initially non-segregated clusters (henceforth NS; i.e., the IMF
is identical throughout the whole cluster) and the other with initially
segregated clusters (henceforth S; i.e. a radially-dependent mass function)
to investigate the effect on the cluster's evolution. Thereafter,
we try models with an initially flattened (bottom-light) IMF to see if these
initial conditions can better reproduce the observed mass function and mass segregation.

To save computational cost, we do not add primordial binaries in
our simulated clusters, although binaries created via three-body
interactions are automatically included. Because of the cluster's
large extent, dynamical effects from primordial binaries are not expected
to be significant \citep{Kroupa95} . Including binaries, however, may decrease the
mass segregation time-scale and increase the star--star collision rate to
some degree.

The evolution of each cluster is followed for Pal~4's estimated
age of 11 Gyr. Stellar and binary evolution is modelled using the
\textsc{sse/bse} routines developed by ~\citet{Hurley00} and  ~\citet{Hurley02}.
Velocity kicks are given to stellar remnants at the time of their formation. Because of the low escape velocity of Pal~4
over most of its lifetime, we (effectively) assume a 0\%
retention fraction for neutron stars and black holes which form
during the simulation, whereas all white dwarfs remain inside
the cluster.


\textsc{nbody6} also includes tidal effects of an analytic
galactic background potential consisting of a bulge, a disc and a
logarithmic halo, which we adjusted to resemble the Milky Way
using the configuration as described in \citet{Allen91}, which yields a rotational velocity of 220 km/s within the disc at 8.5 kpc from the galactic centre. The Allen \& Santillan model has been widely used for the purpose of numerical orbit calculations, owing to a mathematically simple and analytically closed form which is preferred as it supports fast computations. Newer models for the potential field of the Milky Way exist \citep{Wilkinson99,Sakamoto03, Irrgang13}, however the differences are not important for this paper since we consider a circular orbit for Pal~4. The differences might be more important for our future models which include orbits with smaller perigalactica. For the sake of simplicity and due to a lack of observational constraints, we chose a circular orbit for our models of Pal~4 around the Milky Way at the cluster's present-day Galactocentric distance of 102.8\,kpc with the orbital velocity of 200 km/s.

Because of the computational cost of $N$-body simulations (it takes about 3 d for one simulation to complete),  it is hard to do statistics for all parameters by repeating several runs for each simulation. Hence,
in order to determine the influence of statistical scatter on our results, we repeat run M60R14 five times varying the random seed for generating the positions and velocities for each star. We will use the resulting uncertainties in the following discussion. We ran about 60 models with different initial conditions, but we just discuss those models that come close to the observed values which are summarized in Table 1.

\section{Results}\label{Sec:Results}

In this section we present the results from our numerical simulations. In order to compare with observations, we have four main observables: the present-day projected half-light radius, $r_{hl}=18.4\pm 1.1$\,pc, the observed mass in the mass range 0.55 -- 0.85\,M${\odot}$, $M_{measured}=5960\pm110\,\mbox{M}{\odot}$ and the present-day global slope of the mass function $\alpha=1.4\pm0.25$ inside projected radius $r=2.26$\,arcmin in the same mass range, and finally the line of sight velocity dispersion of bright stars of $\sigma=0.87\pm0.18$\,kms$^{-1}$ measured within the tidal radius.
We also compare the final total mass of the modelled clusters with the present-day total mass,
$M_{tot}=29800\pm800\,\mbox{M}{\odot}$, which is obtained by extrapolating the measured mass function towards
lower-mass stars including the contribution of compact remnants \citep{Frank12}.
Due to the low escape velocity of Pal 4 over most of its lifetime, we (effectively) assume a 0\% retention fraction for neutron stars and black holes which form during the simulation, whereas all white dwarfs remain inside the cluster. This is in accordance with the cluster mass estimate of \citet{Frank12}.

Note that in order to match $r_{hl}$ in the $N$-body models we use only
the giant stars, as the light profile is dominated by those stars.  Giant stars are identified in the models by their stellar-evolution phase as red giant and asymptotic giant branch stars.
Moreover, to be consistent with the observations, we restrict our
analysis of the mass function to stars which lie at a projected
distance of less than $r=2.26$\,arcmin ($\approx67$\,pc).
Following Frank et al. (2012), we extract the mass function for stars within this radius with a mass in the range of
$0.55 < m < 0.85\,\mbox{M}{\odot}$. That is, we
ignore compact remnants for this measurement as they are too
faint to be observable in Pal~4. Regarding the velocity
dispersion, we use only stars within the cluster's tidal radius, 
with a mass higher than 0.8\,M${\odot}$.

In addition to these four basic properties, the slope of the mass function in different
radial intervals from the cluster centre out to the projected radius of $r=2.26$\,arcmin ($\approx67$\,pc) is numerically
determined. Adopting different  scenarios for the IMF, we evaluate whether the initially segregated and
unsegregated cluster models can reproduce the observed level of
mass segregation at the current age of Pal~4  (11 Gyr)
as well as its global mass function slope.

In the next sections we present the results of three sets of
simulations:
\begin{enumerate}
\item Kroupa IMF without primordial mass segregation (\emph{canonical-NS}, Sec.~\ref{sec:regular}).
\item Kroupa IMF with different degrees of primordial mass segregation (\emph{canonical-S}, Sec.~\ref{sec:segregated}).
\item Flattened IMF (\emph{flattened}, Sec.~\ref{sec:flattened}).
\end{enumerate}

In Table \ref{tab_regular}, we summarize the initial cluster
properties and the key results of the simulated  clusters, that
is, final radius, total mass, slope of the global mass function and
velocity dispersion.

\subsection{Canonical IMF without primordial mass segregation (canonical-NS)}\label{sec:regular}

\begin{figure}
\includegraphics[width=85mm]{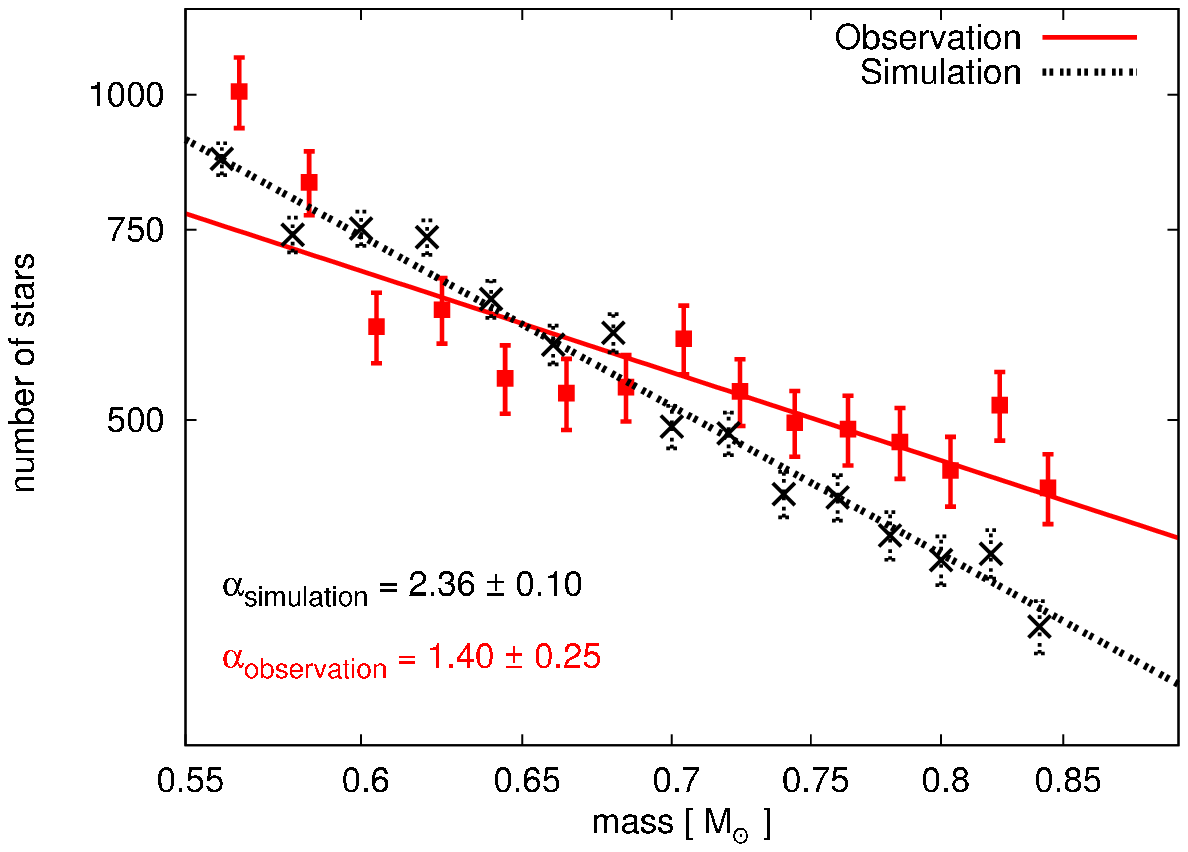}
\caption{Global mass function in the mass range 0.55--0.85\,M${\odot}$ for  model 'M57R14.5' without primordial mass segregation, which started with M=57000 M${\odot}$ and
$R=$14.5\,pc. The mass function at the start of the
simulation was chosen to be a canonical Kroupa IMF, with
$\alpha=2.3$ for the high-mass stars (see Sec.~\ref{sec:regular}).  The red solid line together with the red data points depicts the observed present-day mass function. The black, dotted
line together with the black data points  with a slope of
$\alpha=2.36\pm 0.1$ shows the clusters mass function after an evolution of 11 Gyr. The error of the slope of mass function is derived from fitting.  It is
significantly steeper than the observed value. Hence, two-body
relaxation is not able to deplete the mass function sufficiently
to reproduce the observations, when starting from a
non-segregated cluster on a circular orbit. } \label{mf_regular}
\end{figure}

\begin{figure}
\includegraphics[width=85mm]{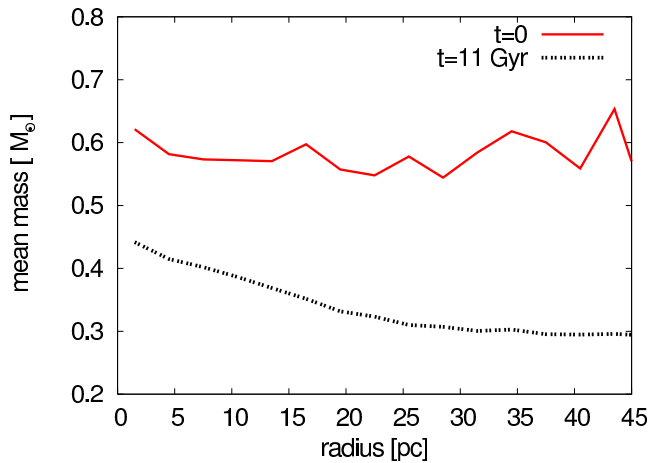}
\caption{Mean stellar mass as a function of 3D radius for model 'M57R14.5' without primordial mass segregation, which started with M=57000 M${\odot}$ and $R=$14.5\,pc, in the mass range 0.55--0.85\,M${\odot}$. Shown are the initial mean mass (red) and the final mean mass at $t=$11 Gyr (black) as function of projected radius.
The decreasing mean mass with increasing distance from the
cluster centre at $t=$11 Gyr  indicates that dynamical mass segregation has happened in the cluster over time. } \label{mean_regular}
\end{figure}

\begin{figure}
\includegraphics[width=84mm]{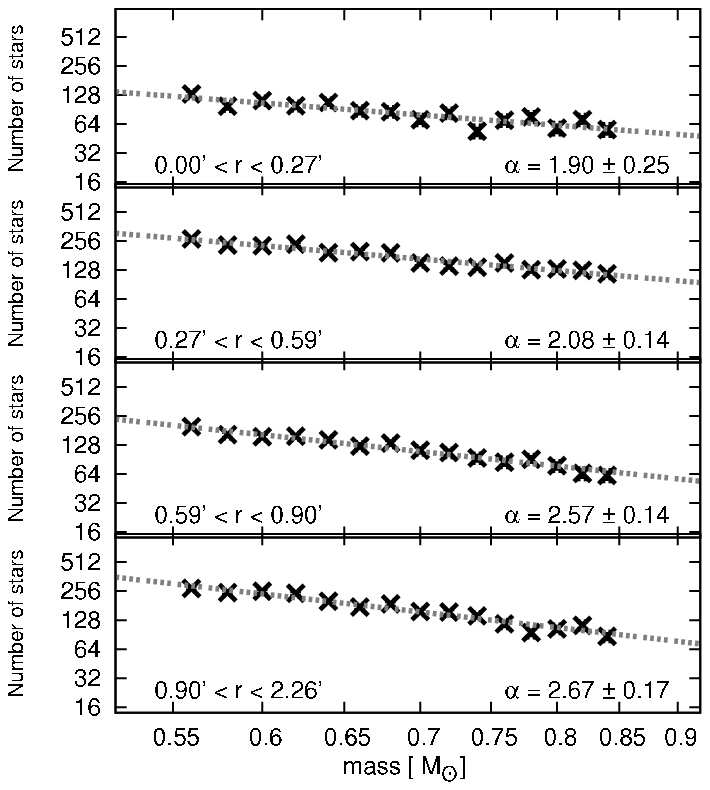}
\includegraphics[width=85mm]{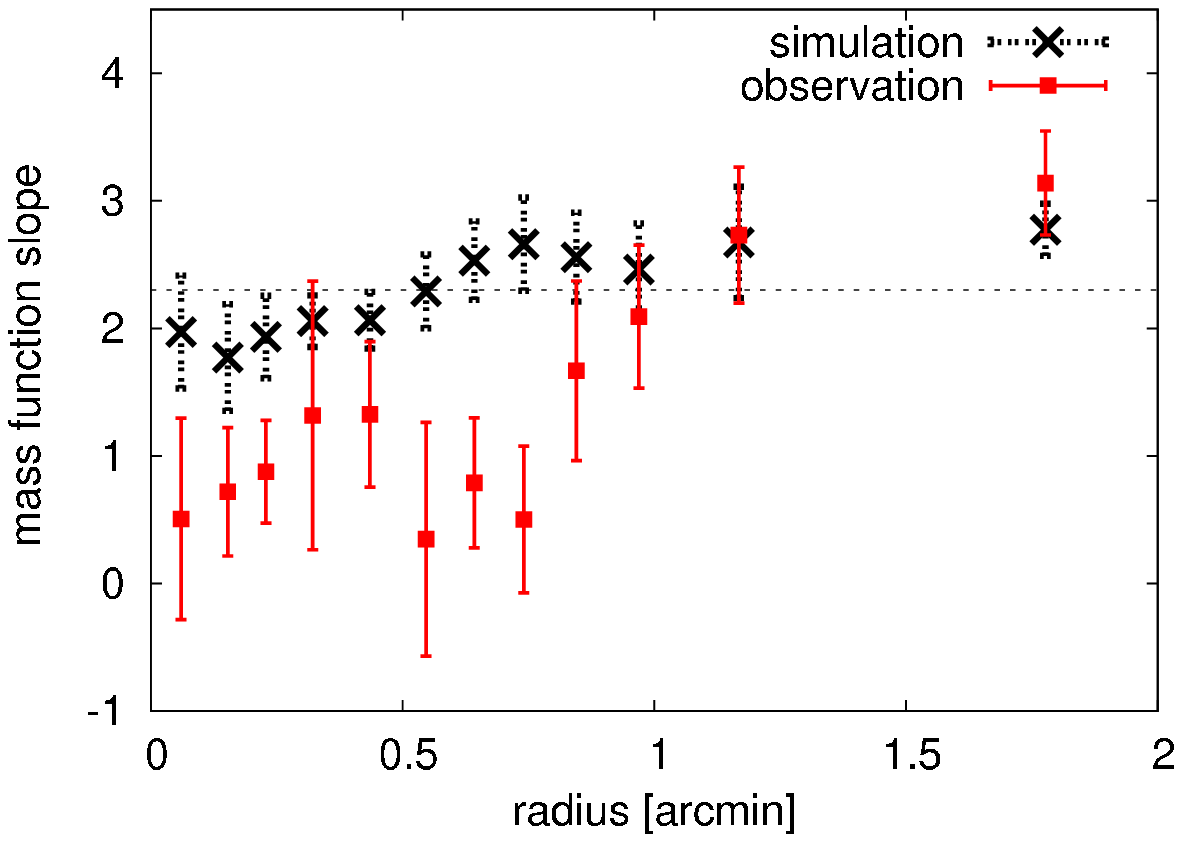}
\caption{ Top: the mass function in the mass range 0.55--0.85\,M${\odot}$  for model 'M57R14.5'  without primordial mass segregation, which started with M=57000 M${\odot}$ and $R=$14.5\,pc,  after an evolution of 11
Gyr in various radial bins. From top to bottom, the panels
represent the innermost to outermost regions shown in fig.~10 of \citet{Frank12}. The black dotted
lines show power-law fits to the data. The radial ranges and
best-fitting power-law slopes are indicated in each panel. The
flattened slope within the inner parts with respect to the slope
in the outer parts indicates that dynamical mass segregation has
happened in the cluster. Bottom: the best-fitting mass function slopes, $\alpha$, derived
in radial bins for the model mentioned above.
The red squares are the observed values taken from
\citet{Frank12}, and the black dots represent the result of the
simulation after an evolution of 11 Gyr.  The $\chi^2$ and $P$-value for this model are 36.24 and 0.001, respectively. This implies a probability of $10^{-3}$ that the data are represented by the model.
 } \label{mfr_regular}
\end{figure}

\begin{table*}
\centering
\caption{Initial and final properties of all simulated
star clusters starting with three different initial mass distributions.
Column~1 gives the model name, in which
the first two digit numbers after `M' denote the initial mass in units of
1000\,M${\odot}$ and the second two digit numbers denote the
initial half-mass radius in parsec. For flattened models the first number in the model name is the adopted high-mass slope (see Section\,\ref{sec:flattened}). Column 2 represents the adopted mass segregation parameter $S$. The mass segregation parameter changes in the range $S=0--0.95$.
The following columns give
the parameters of the simulated star clusters after 11 Gyr of
evolution. The initial two-body relaxation time of models is presented in column 3. Column 4 gives the projected half-light radius $R_{hl}$. Column 5 gives the present-day total mass of the modelled cluster  in the stellar mass range $0.55\leq m/\mbox{M}{\odot} \leq0.85$. Column 6 is the  present-day total mass of the cluster. The present-day slope of the  mass function, $\alpha_{tot}$, for stars with masses between 0.55 and 0.85$\mbox{M}_{\odot}$ inside the projected radius $r=2.26$\,arcmin, corresponding to $\approx67$\,pc, is presented in column 7. Columns 8 and 9 give the slope of the mass function inside and outside the half-light radius, respectively.
The $\chi^2$ goodness-of-fit and corresponding $P$-values are presented in column 10 to compare the observed values of the slope of the mass function in different radial bins to those of the simulated models.
The last column is the line-of-sight velocity dispersion of bright stars.
Compared to the observational values given at the bottom, only the flattened models are an acceptable fit to Pal~4. Typical errors of the numerical values were obtained by repeating run M60R14 five times and are indicated in the header.
The best-fitting models that agree within the uncertainties with all observational
parameters are highlighted with boldface.
}

\begin{tabular}{cccccccccccccc}

\hline (1)&(2)&(3)&(4)&(5)&(6)&(7)&(8)&(9)&(10)&(11)\\
\hline
Model &$S$&$T_{rh}$&$R_{hl}$&$M_{measured}$ &$M^{f}_{r<R_t}$&$\alpha_{tot}$&$\alpha_{in}$&$\alpha_{out}$&$\chi^2$($P$-value)&$\sigma_{los}$\\
& & (Gyr) & (pc)&(M${\odot}$)&(M${\odot}$)&&&&&(kms$^{-1}$)\\
& & &($\pm$2.1)&($\pm$105)&($\pm$1100)&($\pm$0.13)&($\pm$0.13)&($\pm$0.13)&&($\pm$0.02)\\

\hline
Canonical-NS &&&&&&&&&\\
 \hline
       M50R12     &0 &3.18 &16.1   &4894&26755 &2.30   &1.93&2.59&37.08 (0.001)&0.81 \\
       M55R12     &0 &3.28 &16.1   &5354&29372 &2.28   &2.11&2.62&36.84 (0.001)&0.84 \\
       M60R12     &0 &3.43 &15.5   &5920&32418 &2.25   &2.01&2.79&35.76 (0.001)&0.86 \\
       M50R14     &0 &4.01 &16.6   &4953&26762 &2.25   &1.87&2.58&32.52 (0.001)&0.76 \\
       M55R14     &0 &4.13 &19.0   &5236&28506 &2.09   &1.93&2.54&22.44	(0.021)&0.76 \\
       M60R14     &0 &4.32 &16.8   &5670&32320 &2.31   &2.17&2.68&32.28	(0.001)&0.84 \\
       M57R14.5   &0 &4.46 &17.8   &5550&30564 &2.36   &2.18&2.76&36.24	(0.001)&0.80 \\

\hline
Canonical-S &&&&&&&&&\\
\hline

      M60R8  &0.80  &1.86 &12.3   &6449&32823  &2.36   &1.98&3.00&40.80 (0.001)&0.87 \\
      M60R8  &0.95  &1.86 &15.8   &6551&32683  &2.17   &1.68&2.91&25.80 (0.007)&0.79 \\
      M55R9  &0.95  &2.13 &20.5   &5565&27974  &2.19   &1.66&2.89&13.31 (0.273)&0.68 \\
      M60R10 &0.50  &2.60 &14.0   &6084&32203  &2.21   &1.94&2.71&35.50 (0.001)&0.87 \\
      M60R10 &0.95  &2.60 &22.8   &6278&31500  &2.14   &1.56&2.97&10.82	(0.460)&0.67 \\
      M57R10 &0.95  &2.55 &22.1   &5912&30025  &2.25   &1.86&2.97&18.96	(0.062)&0.67 \\
\hline
Flattened   &&&&&&&&&\\
\hline
      F0.5M57R14  &0   &4.23 &20.5   &5295&29435  &1.26   &1.12&1.81&16.87 (0.111)&0.79 \\
      F0.6M57R14  &0   &4.23 &19.6   &5870&30820  &1.56   &1.45&2.04&16.85 (0.112)&0.81 \\
      F0.7M57R14  &0   &4.23 &18.6   &6240&32292  &1.76   &1.60&2.27&20.76 (0.036)&0.84 \\
      F0.6M57R10  &0.50&2.55 &15.5   &5888&30484  &1.51   &1.40&2.06&16.80 (0.114)&0.86 \\
      \textbf{F0.6M55R10}  &0.70&2.49 &16.8   &5746&29418  &1.54   &1.23&2.22&8.88 ~(0.640)&0.80 \\
      \textbf{F0.6M57R12}  &0.70&3.36 &20.5   &5875&30457  &1.61   &1.38&2.28&9.01 ~(0.620)&0.77 \\
      \textbf{F0.6M57R10}  &0.90&2.55 &20.3   &5715&30231  &1.41   &0.79&2.35&13.80  (0.240)&0.76 \\

      \hline
    Observations   & &  &18.4$\pm$1.1 & 5960$\pm$110& 29800$\pm$800  & 1.4$\pm$0.25 &0.89$\pm$0.39&1.87$\pm$0.24&& 0.87$\pm$0.18    \\
       \hline
\end{tabular}
\label{tab_regular}
\end{table*}

All clusters in the first part of Table \ref{tab_regular} start with a
canonical Kroupa mass function but different initial half-mass
radii and initial cluster masses.

As can be seen from the upper section of Table~\ref{tab_regular}, the line-of-sight
velocity dispersion in most of our computed models is about
$0.8$\,kms$^{-1}$,  and therefore within the uncertainties of the
observational value.

Furthermore, many models can reproduce Pal~14's half-light radius
(column 4) and its measured and total mass (column 5 and 6). We also calculate the global mass function slope for all models (column 7). Fig.~\ref{mf_regular} depicts the mass function of one example model after 11 Gyr of evolution and compares it with the
observed data.  Our simulations have typically
$\alpha\simeq2.2$ after 11 Gyr, which means only a mild decrease
of the high-mass slope from the initial Kroupa value of
$\alpha=2.3$. In contrast, the observed slope is $\alpha=1.4 \pm
0.25$. Hence, the models starting with a canonical mass function are unable to reproduce
the observed slope of the mass function of Pal~4 even when accounting for and statistical errors. This is a result of the long two-body relaxation time of the models and of the present-day cluster. This implies an evaporation time-scale which is significantly larger than the age of the cluster.

In Fig.~\ref{mean_regular}, we plot
the mean stellar mass, including the remnants, as a function of
radius for the cluster M57R14.5, which is closest to Pal~4
among our non-segregated clusters.
After 11 Gyr of evolution, the mean stellar mass decreases from $<$m$>\simeq0.45 M{\odot}$ at the centre of the cluster to $<$m$>\simeq 0.30 M{\odot}$ at the half-mass radius. The decline of the mean mass with cluster radius shows the substantial degree of mass segregation that has been generated by dynamical evolution.

In order to test for mass segregation, we also measure the mass function as a function of radius of the modelled cluster. Fig. \ref{mfr_regular} shows the exponents of the best-fitting power laws fitted to the projected mass functions in four different radial bins. After 11 Gyr of evolution, the mass function steepens with increasing radius, from $\alpha \simeq 1.9 \pm 0.25$ within a projected radius of $R=8.1$\,pc to $\alpha \simeq 2.67\pm 0.17$ at the largest observed radius.

Fig.~\ref{mfr_regular} shows that, while the mass function in the  outer region of the non-segregated cluster is almost in agreement with the
observations, this is not the case for the inner part of the cluster.

In order to measure the quality of the fit of the model to the observations, we employ the  $\chi^2$ goodness-of-fit test defined as:

\begin{equation}
 \chi^2=\sum_{i=1}^{N}\frac{(\alpha_{i, sim}-\alpha_{i, obs})^2}{\sigma_i^2}\label{chi},
\end{equation}
where $\sigma_i^2=\sigma_{i, sim}^2 + \sigma_{i, obs}^2$ is the uncertainty in the slope of the mass function including both observational as
well as the simulated uncertainty. The sum runs over the $N=12$ radial bins.
We obtain a value of $\chi^2=36.24$, which shows that, although the simulations show some degree of mass segregation, the primordially unsegregated clusters do not attain the observed degree of mass segregation and the discrepancy with the observations is extremely large. The corresponding $P$-value of 0.001 allows us to reject this model at the 99.9\%
confidence level. Similar results for the slope of the mass function are obtained for clusters with other initial radii and masses. These models do not undergo much mass loss due to being on circular orbits at a large Galactocentric radius, so their mass function can not fit after 11 Gyr of dynamical evolution. The main difference among
these models lies in their final masses and radii. We therefore conclude that two-body relaxation alone cannot be responsible for the flattened mass function of Pal~4 and its segregated structure, if the post-gas expulsion re-virialized cluster had a canonical Kroupa IMF.

\subsection{Canonical IMF with primordial mass segregation (canonical-S)}\label{sec:segregated}

\begin{figure}
\includegraphics[width=85 mm]{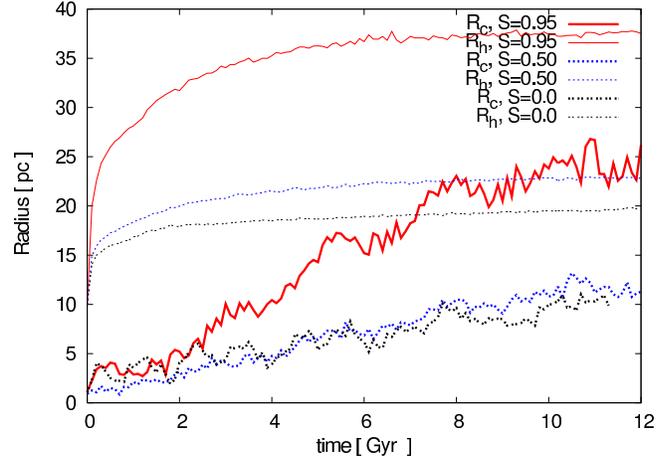}
\caption{ Evolution of the 3D half-mass radius (upper thick curves) and core radius (thin curves) for different degrees
of primordial mass segregation with a canonical Kroupa IMF. The
initial mass is 57000 M${\odot}$, and the initial half-mass radius
is 10 pc for all models.
Because of stellar evolution clusters experience a rapid expansion such
that clusters with higher degrees of primordial mass segregation
experience a larger jump in the half-mass radius within the first
100 Myr of evolution. The rising core radius shows that the cluster is still in the pre-core-collapse phase. }
\label{radius}
\end{figure}

\begin{figure}
\includegraphics[width=85mm]{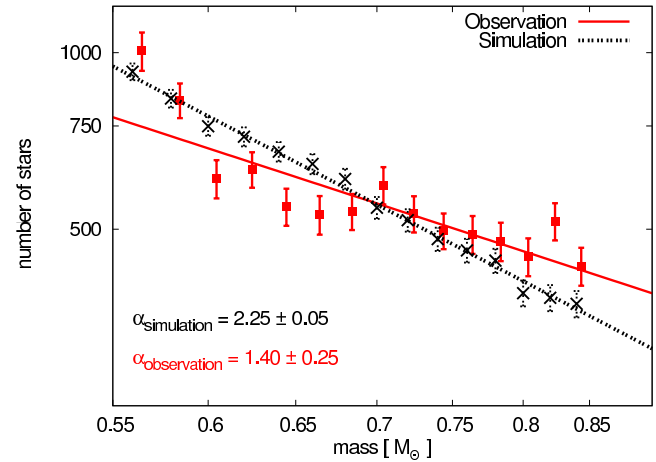}
\caption{The same as Fig. \ref{mf_regular}, but for a primordially mass-segregated cluster with a canonical
Kroupa IMF. The initial mass of this particular model
(M57R10) is 57000 M${\odot}$, the initial half-light radius
is 10 pc, and the mass segregation parameter is set to $S=0.95$.
Clusters with such a strong degree of primordial mass segregation
are still not able to reproduce the observed flat mass function (see the text for more explanation).}
\label{mf_seg}
\end{figure}

\begin{figure}
\includegraphics[width=85 mm]{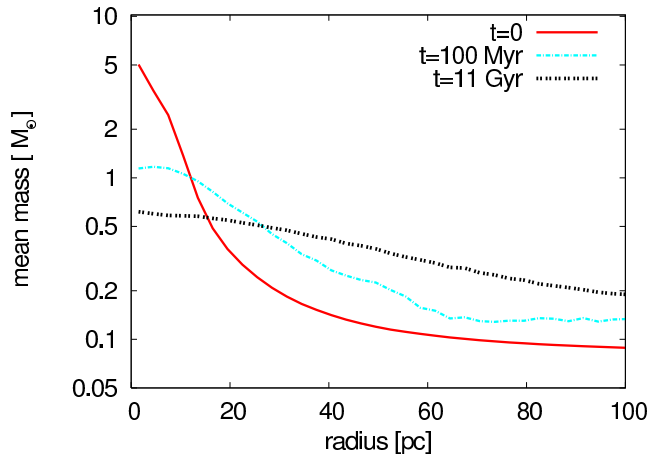}
\caption{The same as Fig. \ref{mean_regular}, but here we
started with a primordially mass-segregated cluster with a
canonical Kroupa IMF. The initial mass of this particular model
(M57R10) is 57000 M${\odot}$, the initial half-light radius
is 10 pc and the mass segregation parameter is set to $S=0.95$. The mean mass profile at $t = 100$ Myr after the
early stellar evolution of massive stars is also plotted.
 } \label{mean_seg}
\end{figure}

\begin{figure}
\includegraphics[width=84mm]{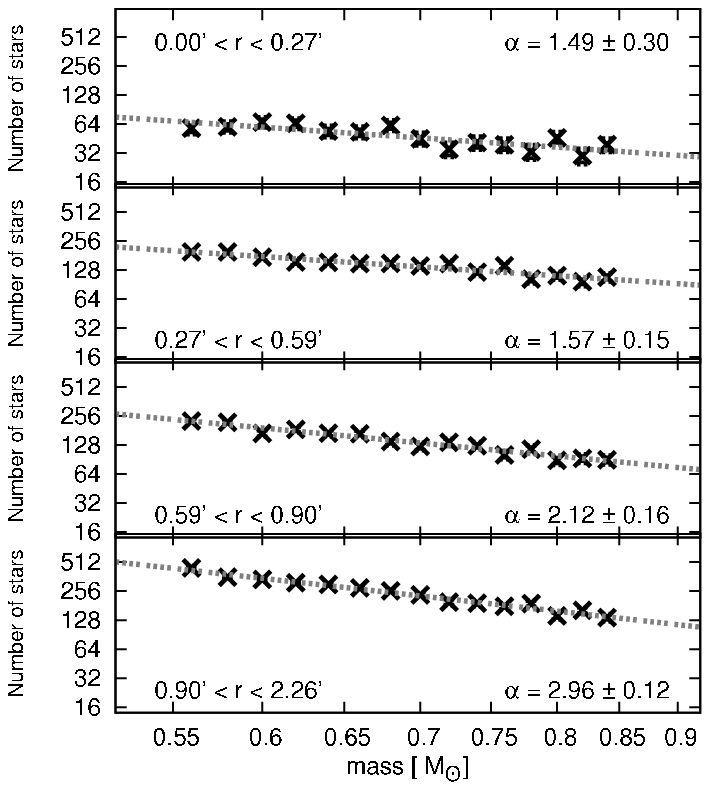}
\includegraphics[width=85mm]{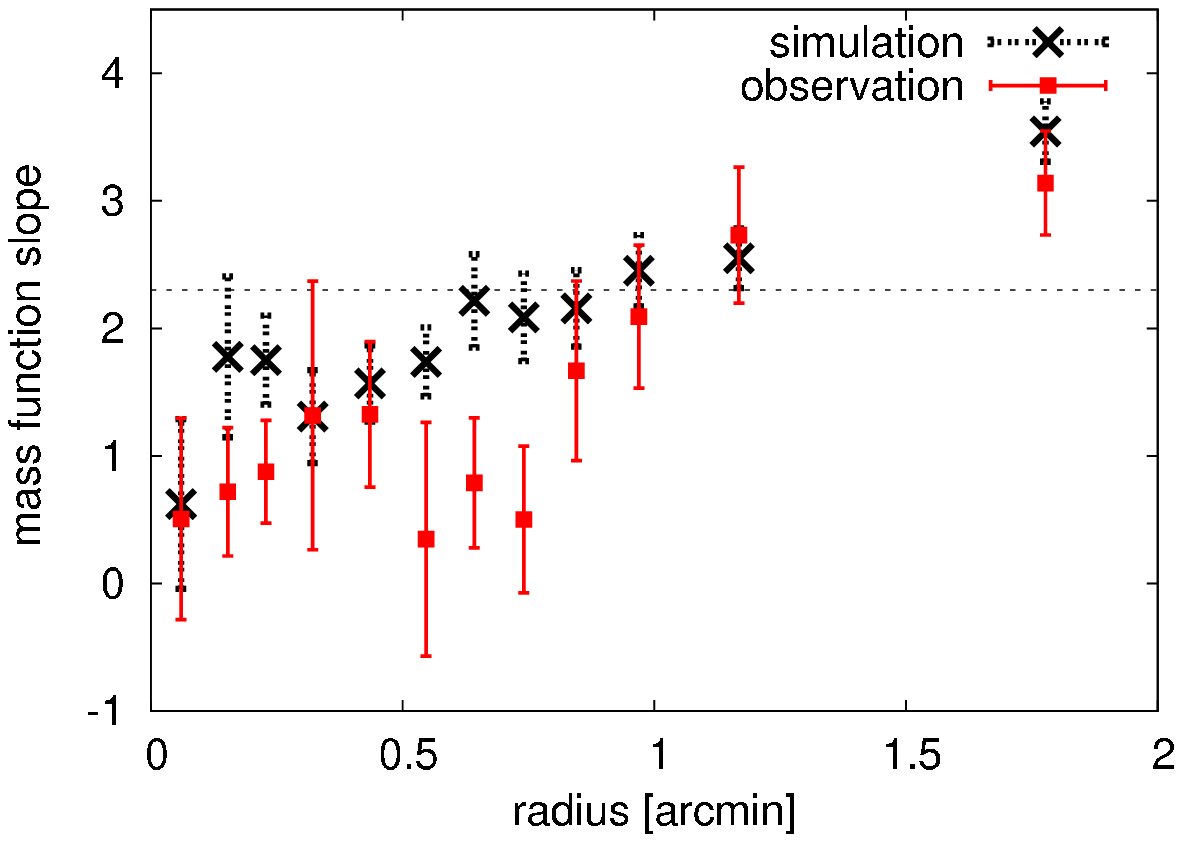}
\caption{ The same as Fig. \ref{mfr_regular}, but for a primordially mass-segregated cluster with a
canonical Kroupa IMF. The initial mass of this particular model
(M57R10) is 57000 M${\odot}$, the initial half-light radius
is 10 pc and the mass segregation parameter is set to $S=0.95$. As can be seen,
the mass function steepens with increasing radius, and the mass function slopes in each bin are marginally compatible with the observed values.
The $\chi^2$ value for this model is 18.96 corresponding to a probability of 6.2\% that the data can be represented by the model.
 } \label{mfr_seg}
\end{figure}

In order to understand whether primordial mass segregation helps to reconcile the inconsistency between observations and simulations, we calculated a number of models starting with primordial mass segregation. That is, the post-gas expulsion re-virialaized cluster is assumed to be mass segregated with an overall canonical IMF.

The code \textsc{McLuster} allows to initialize  any degree of primordial mass segregation (hereafter, $S$) to all available
density profiles.  $S=0$ means no segregation ('NS' in Table 1) and $S=1$ refers to full segregation. \textsc{McLuster} uses the routine described in
\citet{Baumgardt08a} to segregate the clusters. This routine allows to maintain the desired density profile when increasing
the degree of mass segregation while also making sure that the cluster is in virial equilibrium. For a fully segregated cluster, the highest mass
star occupies the orbit with the lowest energy.

To allow for a better comparison with the models  from Sec.~\ref{sec:regular}, we set up clusters with various degrees of mass segregation in the range $S=0.5-0.95$, the same range of initial cluster masses, but with smaller initial half-mass radii. The reason that we chose a smaller initial radius is the expansion due to both dynamical and stellar evolution. In the first few Myr of a cluster's life, the mass loss is dominated by stellar evolution of massive stars
in the core and leads to a rapid expansion in the size of the cluster. Fig.~\ref{radius} shows the evolution of the 3D half-mass radius and the core radius for different values of the mass segregation parameter over the entire evolution of the cluster.  The growing core radius shows that the cluster is far from the core collapse.

The results of the simulated models with primordial mass segregation are shown in Table~\ref{tab_regular} and Figs.~\ref{mf_seg}--\ref{mfr_seg}. Even by selecting very high values of primordial segregation, the present-day global mass function of Pal 4 cannot be reproduced.
Since the whole cluster is included within 2.26\,arcmin in the calculation of the global mass function, and since the overall mass function did not change for the segregated clusters (because we just distributed the stars differently according to their mass), one should expect to end with the same global (i.e. within $R=2.26$\,arcmin) mass function as in the unsegregated case. Also, distant globular clusters undergo very little mass loss due to tidal interaction. Almost 45\% of the initial mass is lost owing to early stellar evolution.  From Table~\ref{tab_regular} it can then be estimated that dynamical interactions lead to about 15\% additional mass loss after 11 Gyr evolution.

Fig.~\ref{mean_seg} shows the mean stellar mass, including the remnants, as a function of 3D radius for the modelled cluster 'M57R10' with $S=0.95$.
The rapid fall in mean mass during the first 100 Myr is due to stellar evolution of massive stars, which is the dominant process in the early
evolution of the clusters. Thereafter, the change in mean mass continues more slowly.  The final mean mass profile at 11 Gyr shows that the mean stellar mass decreases from $<$m$>\simeq0.62 M{\odot}$ at the centre of the cluster to $<$m$>\simeq 0.43 M{\odot}$ at the 3D half-mass radius of 37 pc.

By comparing Figs ~\ref{mfr_regular} and~\ref{mfr_seg}, it can be seen that the mass function slopes in different radial bins are closer to the observed values for the initially segregated models than the unsegregated ones. This is confirmed by the lower $\chi^2$ value of 18.96 for this model. The corresponding $P$-value of 0.062 allows us to reject this model at the 93\% confidence. Moreover,  in the segregated models, the amount of mass function flattening does not depend on the amount of mass segregation.

According to Table~\ref{tab_regular},  some of the mass segregated models such as  'M55R9', 'M60R10', and 'M57R10' have $P$-values of 0.273, 0.460 and 0.062, respectively, so these models cannot be excluded given the data of radial structure of the slope of the mass function, but the difference between the observed and calculated values of $\alpha_{tot}$ are significant for both inside and outside the half-light radius. The slope of
the mass function in all models is far from the observed value of Pal 4, though, even when accounting for observational and statistical errors. Moreover, we find that the values for the present-day total mass of the modelled clusters M55R9' and 'M60R10' in the stellar mass range  $0.55\leq m/\mbox{M}{\odot} \leq0.85$ (column 5 of Table 1) are not compatible with the observed values. A glimpse on the last column of Table 1 shows that if we account for the statistical and observational errors, the line-of-sight velocity dispersion for these models is marginally compatible with the observed value. Therefore it is clear that we will not be able to fit the observations if we start with a Kroupa/Salpeter mass function.

\subsection{Flattened IMF}\label{sec:flattened}

\begin{figure}
\includegraphics[width=85mm]{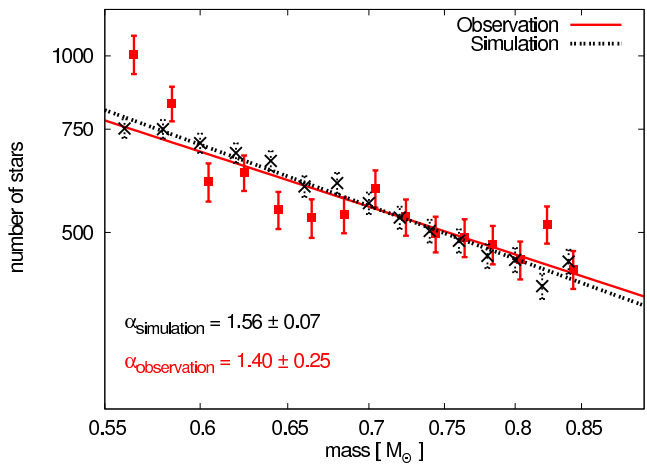}
\caption{The same as Fig. \ref{mf_regular}, but here we started
with a flattened mass function with a slope of $\alpha=1.6$ above 0.5 M${\odot}$ and $\alpha=0.6$ below 0.5 M${\odot}$. The initial mass of this particular model
(F0.6M57R14) is 57000 M${\odot}$, the initial half-light radius
is 14 pc.  Further properties of the cluster after 11 Gyr of
evolution are given in Table~\ref{tab_regular}.}
\label{mf_flat}
\end{figure}

\begin{figure}
\includegraphics[width=84 mm]{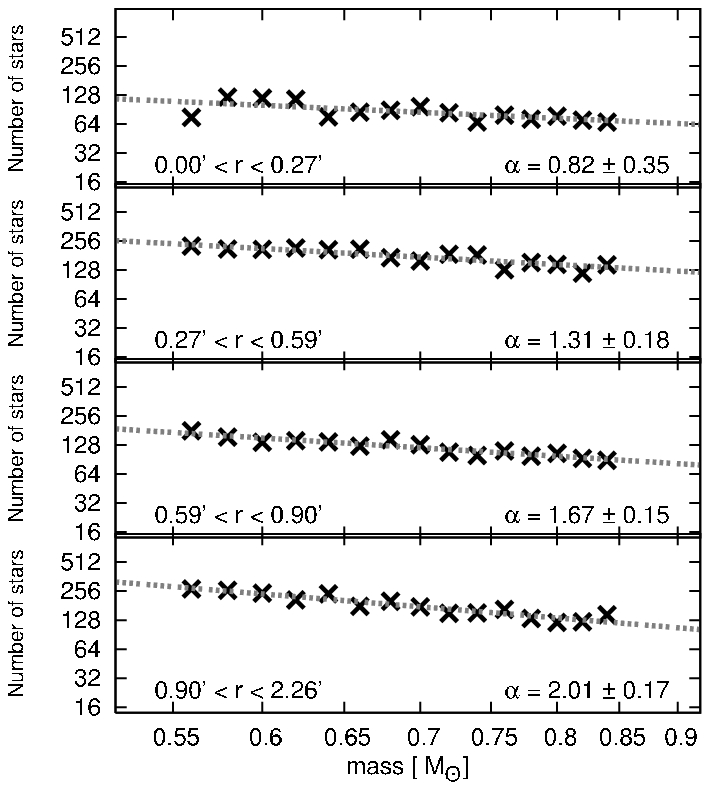}
\includegraphics[width=85 mm]{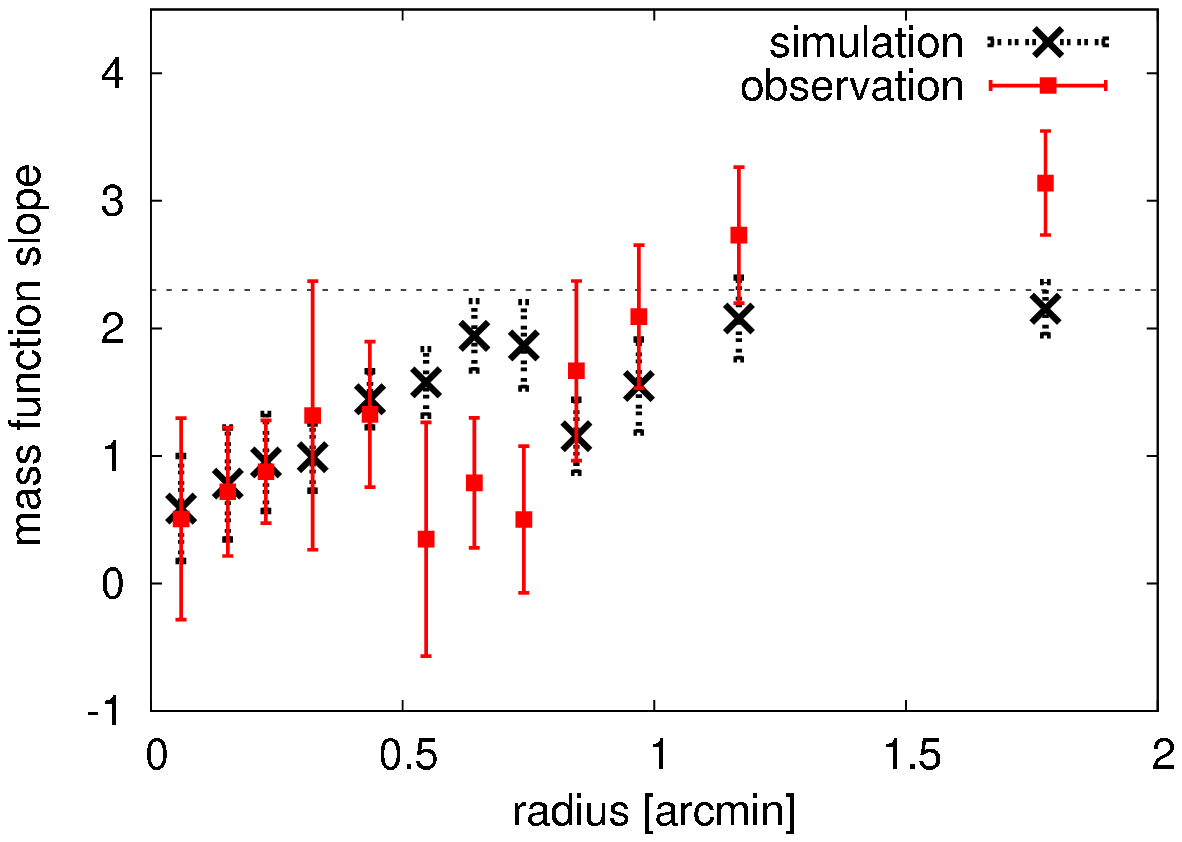}
\caption{The same as Figure \ref{mfr_regular}, but here we started
with a flattened mass function with a slope of $\alpha=1.6$ above 0.5 M${\odot}$ and $\alpha=0.6$ below 0.5 M${\odot}$. The initial mass of this particular model (F0.6M57R14) is 57000 M${\odot}$, the initial half-light radius
is 14 pc. Further properties of the cluster after 11 Gyr of
evolution are given in Table~\ref{tab_regular}. The $\chi^2$ value for this model is 16.80, corresponding to a probability of 88.6\% for rejecting the model.
 } \label{mfr_flat}
\end{figure}

So far, we have shown that clusters starting with a Kroupa/Salpeter slope of $\alpha=2.3$ at the high-mass end cannot reproduce the observed slope of the mass function well. The smallest slope that can be achieved from
these models is about 2.0, while the observed slope is
$\alpha=1.40\pm0.25$. In this section we present a number of models starting with a flatter IMF to see whether it is possible to make up for the discrepancy in $\alpha$.

One way to achieve such a flat mass function is if the cluster was born mass segregated and
embedded in a primordial gas cloud. In such a case, the early phase of gas expulsion can be very violent, depending on the exact initial conditions of its parent gas cloud \citep{Baumgardt07,
Baumgardt08b, Parmentier08, Marks08, Dabringhausen10, Banerjee13}. The ejection and distribution of the remaining cloud gas is the natural outcome of the stellar winds of massive stars, and of supernovae explosions. This early rapid mass loss changes the gravitational potential, and consequently causes a cluster
to expand. Such expansion leads to the rapid dissolution of low-concentration clusters. For initially mass-segregated
clusters, the mass lost due to the evolution of massive stars is removed preferentially from the cluster's inner regions, and the
early expansion of the cluster is stronger and potentially more destructive than when the same amount of mass is lost in a
non-segregated cluster.

\cite{Marks08} suggested that an initially mass-segregated cluster loses preferentially
low-mass stars during the gas-expulsion phase, which would leave the cluster with a flattened mass function.
Therefore, a globular cluster's evolution over a Hubble time can be strongly affected by this early evolutionary processes \citep{Marks10}.

Assuming that a certain flattening of the mass function slope has happened within the first
100 Myr of the cluster's evolution, we have performed a series of $N$-body
simulations for models with various flattened initial slopes of the
mass function instead of the canonical IMF (Table \ref{tab_regular}).
The effect of early evolutionary processes on the long-term evolution of clusters cannot be easily computed in
a direct way for a globular cluster of the pre-gas expulsive mass of Pal 4, because clusters can lose a large
fraction of their birth stellar population as a result of gas expulsion.
The initial state of our models therefore ought to be understood as being the re-virialized state of a post-gas expulsion cluster \citep{Banerjee13}.

Since stars with masses larger than 5 M${\odot}$ will have died or turned into
compact remnants within the first 100 Myr, the maximum mass in the mass spectrum was set to 5 M${\odot}$,
instead of 100 M${\odot}$. Because of the low escape velocity from Pal 4, it is reasonable to assume a 0 per
cent retention fraction for neutron stars and black holes that form
during the simulation.

We have chosen three sets of slopes for the
mass function: $\alpha_a=\{1.7,0.7\}$, $\alpha_b=\{1.6,0.6\}$ and
$\alpha_c=\{1.5,0.5\}$, where the first number in each set is the
slope of the mass function for stars more massive than 0.5
M${\odot}$ and the second one is for low-mass stars (for
comparison, the Kroupa IMF is $\alpha = \{2.3,1.3\}$). The first
column in Table \ref{tab_regular} shows the name of the simulated
models. For example, F0.6M57R14 represents a flattened model with
a slope of $\alpha_b=\{1.6,0.6\}$, an initial mass of 57000 M${\odot}$, and
an initial (i.e. post-gas expulsion re-virialized) half-mass radius of 14 pc.

The results are summarized in the lower section of Table \ref{tab_regular}. We find that a
particular model (F0.6M57R14) without primordial segregation ($S=0$)  can reproduce the present
day total mass and half-light radius of Pal 4 better than the scenarios mentioned above.
The global mass function slope of this model after 11 Gyr of evolution is  $\alpha=1.56\pm0.07$, which is
compatible with the observed value within the uncertainty (see also Fig. \ref{mf_flat}).
Moreover, according to Table \ref{tab_regular}, all other present day properties of this model are
in good agreement with the observed values.

In Fig. \ref{mfr_flat} we plot the mass function slope as a function of radius for the best-fitting model, F0.6M57R14. We see that mass segregation has taken place, in addition to our artificial flattening of the mass function, by about the same amount as in the non-flattened case (Fig.~\ref{mfr_regular}).

Fig. \ref{mfr_flat} shows the mass function
and power-law fits to the mass distributions for different radii.
As can be seen, the slope of the inner part is compatible with the observed value.  The average values of the mass function slopes in the
outer regions are compatible with the observed values within
the uncertainties, since we obtain $\chi^2 = 16.8$, which corresponds
to a $P$-value of 0.114. This means that the model can be just rejected with 88.6 per cent confidence. That is not very strong and this model can be acceptable.

\begin{figure}
\includegraphics[width=85mm]{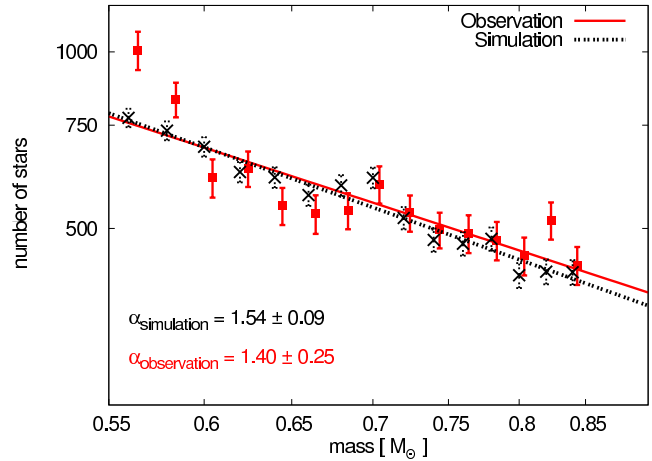}
\caption{The same as Fig. \ref{mf_regular}, but here we started
with a primordially segregated cluster with a flattened mass function with a slope of $\alpha=1.6$ above 0.5 M${\odot}$ and $\alpha=0.6$ below 0.5 M${\odot}$ . The initial mass of this particular model
(F0.6M55R10) is 55000 M${\odot}$, the initial half-light radius
is 10 pc, and the mass segregation parameter is set to $S=0.70$.  Further properties of the cluster after 11 Gyr of
evolution are given in Table~\ref{tab_regular}.}
\label{mf_seg_flat}
\end{figure}

\begin{figure}
\includegraphics[width=84 mm]{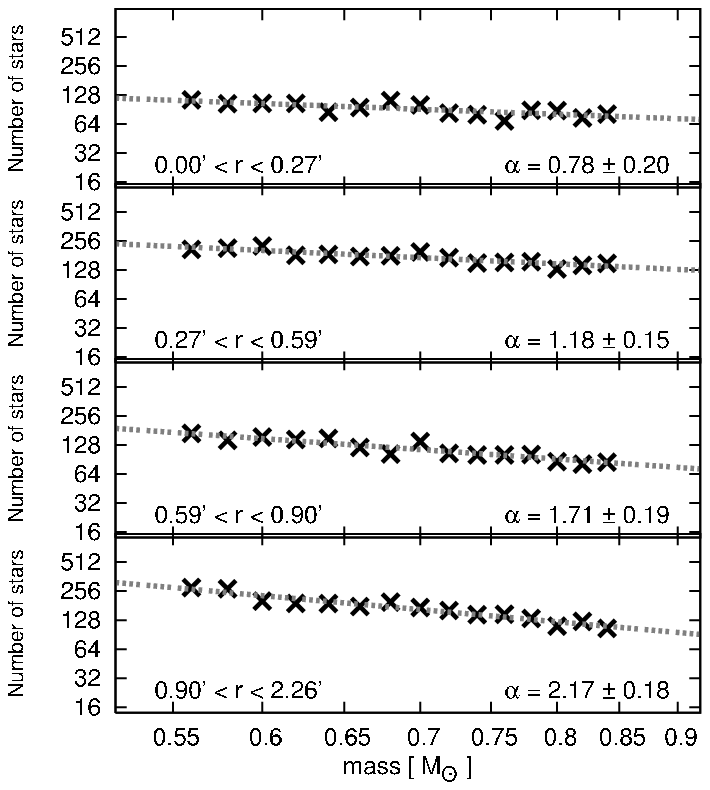}
\includegraphics[width=85 mm]{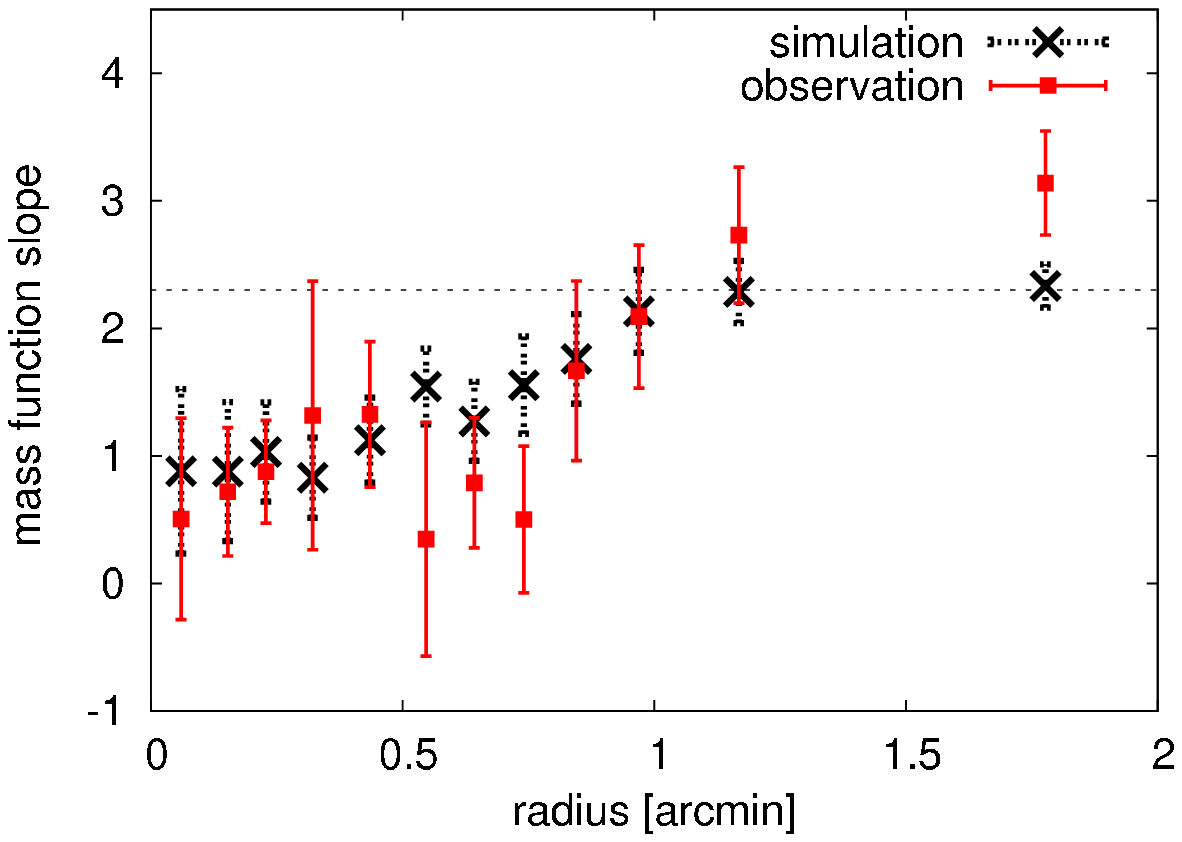}
\caption{The same as Fig. \ref{mfr_regular}, but here we started
with a flattened mass function with a slope of $\alpha=1.6$ above 0.5 M${\odot}$ and $\alpha=0.6$ below 0.5 M${\odot}$. The initial mass of this particular model
(F0.6M55R10) is 55000 M${\odot}$, the initial half-light radius
is 10 pc and the mass segregation parameter is set to $S=0.70$. Further properties of the cluster after 11 Gyr of
evolution are given in Table~\ref{tab_regular}. The  $\chi^2$ value for this model is 8.9, corresponding to a $P$-value of 0.64.
 } \label{mfr_seg_flat}
\end{figure}

This agreement can be increased by adding mass segregation to the initial configuration. Such a cluster with a flattened mass function and some left-over mass segregation ($S\geq0.7$) would be the natural outcome of a primordially strongly mass-segregated cluster which expelled its gas and preferentially lost its low-mass stars. As shown in Figs~\ref{mf_seg_flat} and \ref{mfr_seg_flat}, the agreement can be made almost perfect, since we obtain $\chi^2=8.9$, which corresponds to a $P$-values of 0.640.

In fact, in models with flattened mass function, the slope of the stellar mass function in the mass range  $0.55\leq m/\mbox{M}{\odot} \leq0.85$  is lowered across the cluster in all radial bins. Because of the addition of mass segregation, the slope of the mass function in the outer radial bins ($r>0.7$\,arcmin) increases, while in the inner bins ($0.7>r>0.3$\,arcmin) it decreases. This leads to a better agreement with the observations.

\subsection{The effect of unresolved binaries}

Binary stars, either primordial or dynamically formed during
close encounters between single stars, can affect the
observational parameters of a star cluster, such as the velocity
dispersion and the mass function. Unfortunately, binaries slow down
direct $N$-body computations enormously because timesteps have
to be very small for their integration, such that most numerical
investigations neglect binaries completely. Star cluster models with a 100\% primordial binary fraction and full realistic binary star distribution function have been presented by Kroupa (1995a, b, c) and \citet{Kroupa01b}. The binaries could also have an impact on the dynamical evolution of the cluster, and might therefore allow the initial conditions to be more compact (allowing rapid mass segregation) followed by a period of binary-driven expansion \citep{Wilkinson03}.

In our analysis we did not take into account the effects of
unresolved binaries on the determination of the mass function
so far. An unresolved binary consisting of two
main-sequence stars will have a combined colour somewhere in
between the colours of the two components, and a magnitude
brighter than that of a single-star main sequence at this
combined colour  \citep{Kroupa92}.
If the binarity is not taken into account in the determination of stellar masses, the combined system will be assigned
a mass that is larger than the mass of the two single stars. This causes an unrealistic flattening in the mass
function slope \citep{Kroupa91,Kroupa93}.

The magnitude of this effect depends on the fraction of unresolved binaries and the mass distribution of the binary components.
Here, we investigate the effect of binaries on measured mass function slopes by populating evolved \textsc{McLuster} models of Pal~4 with a varying fraction of binaries,
\begin{equation}
f_{bin}=\frac{N_{bin}}{(N_{bin}+N_{single})},
\end{equation}
following the method outlined in \citet{Frank12}.  $N_{bin}$ is the number of binary systems in the cluster
and $N_{single}$ is the number of single stars. We add binaries following a Kroupa period
distribution and a thermal eccentricity distribution (\citealt{Kroupa95b}), and the binary components with different masses were paired randomly for stars with $ m\leq 5 M{\odot}$ after choosing both companion masses independently and randomly from the IMF. For massive stars with $ m\geq 5 M{\odot}$ the pairing rules change and they tend to prefer more similar-mass companions \citep{Sana11}.

We assume different values for the binary fraction ranging from 10 to 90 per cent, to evaluate the effect of a population of unresolved binaries on the slope of the mass function.
To obtain the mass function of the cluster members, we do the following procedure.

(i) We derive the luminosity of each star in the modelled cluster and add up the luminosities of the two components for each binary system.

(ii) In order to turn the combined luminosity back into a mass, we generate a large cluster model, astrophysically evolve it to an age of 11 Gyr  and derive a relation between luminosity and mass from all single stars in the cluster. Using this relation, we then convert the luminosity of each binary back into a mass estimate.

It should be noted that, in order to be compatible with the observations for Pal~4 by Frank et al. (2012),  we restrict the sample from which we derive the transformation to systems for which at least one component is a main-sequence, sub-giant branch, or red giant branch star.

\begin{figure}
\begin{center}
\includegraphics[width=85 mm]{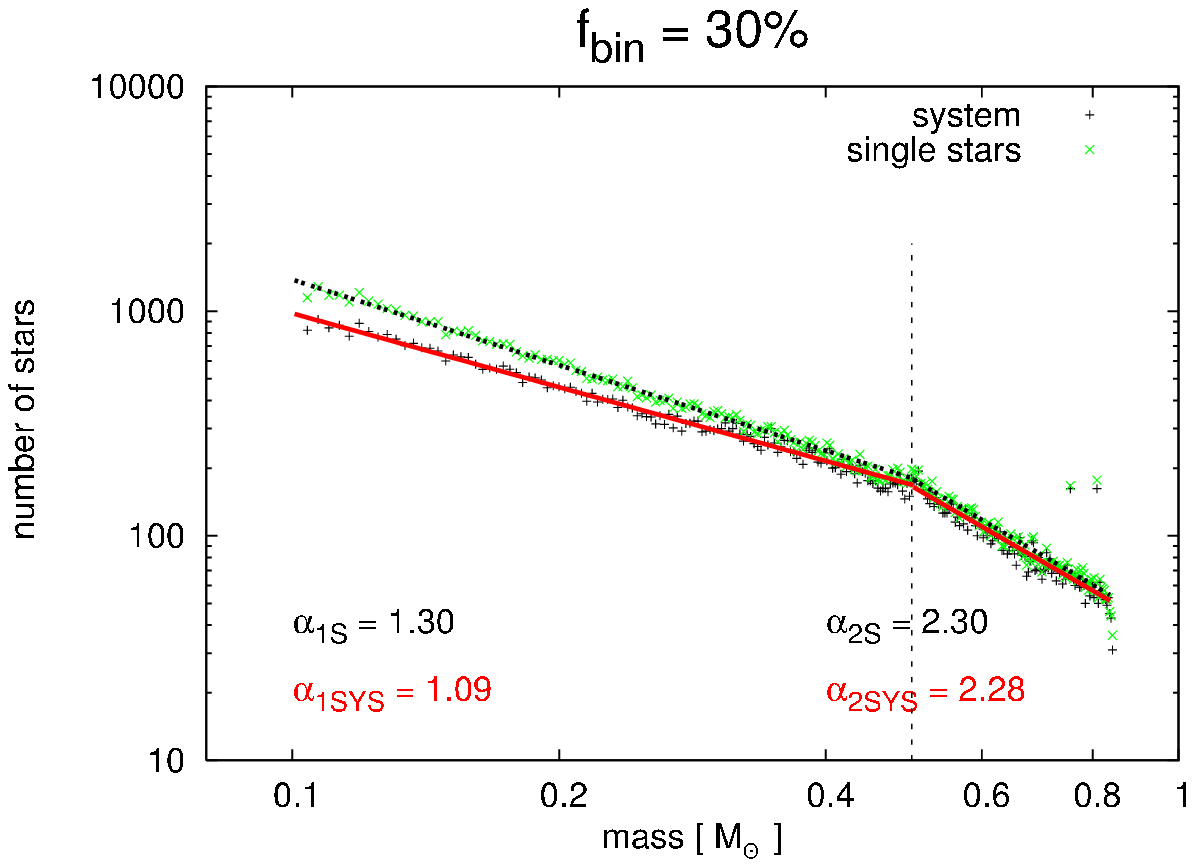}
\includegraphics[width=85 mm]{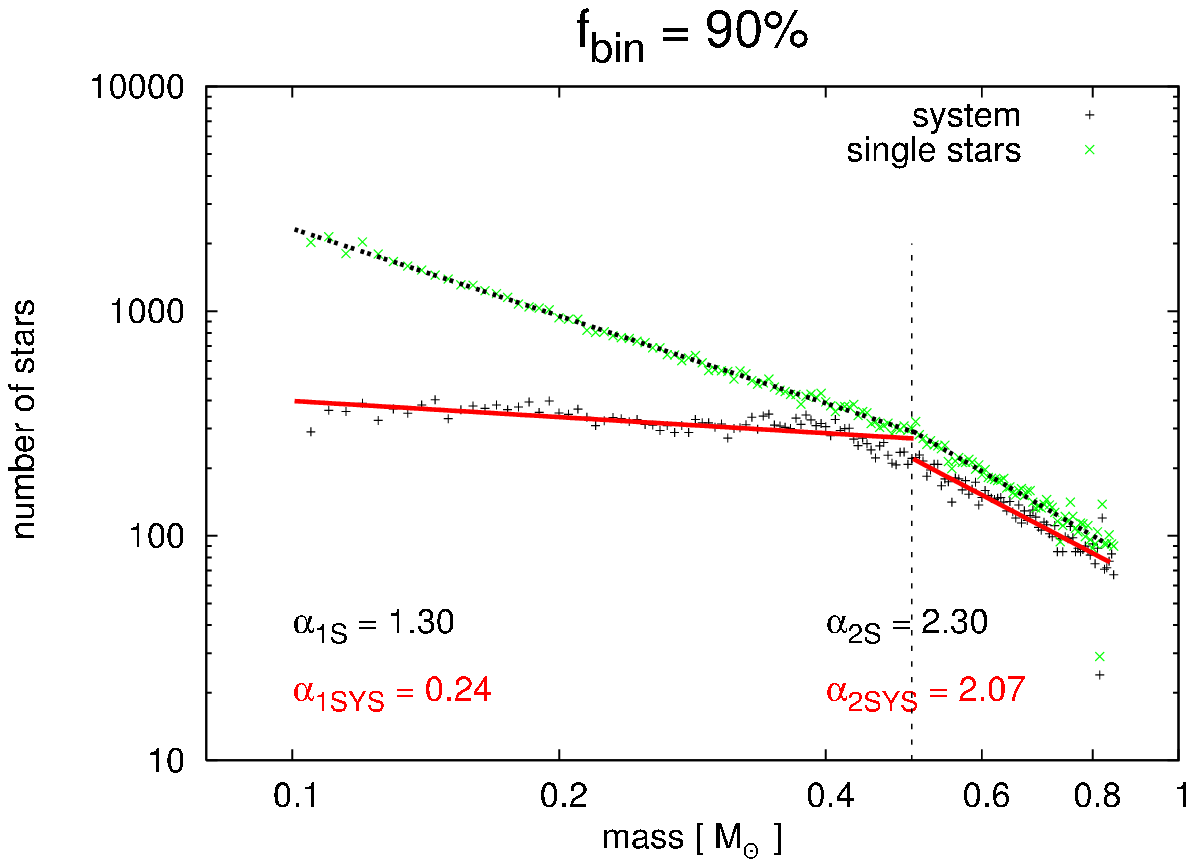}
\caption{The effect of increasing the binarity on the measured mass function. In the top panel, the binary fraction is  $f_{bin}=30\%$ and in the bottom panel it is $f_{bin}=90\%$. The vertical dashed line which corresponds to $m_{break}=0.5 M{\odot}$ shows the point at which the Kroupa canonical IMF has a break. The black, dotted lines show the best fit to the mass function of low- and high-mass stars assuming all binaries to be resolved into their components. The corresponding values for the slopes ($\alpha_{1S}$ and $\alpha_{2S}$) in both low- and high-mass ranges are compatible with a Kroupa IMF. The solid red lines show the best fit to the mass function considering the effect of unresolved binaries. According to the calculated values for the slopes ($\alpha_{1SYS}$ and $\alpha_{2SYS}$), a high binary fraction can severely affect the measured mass function slope in the low-mass range, below $m=0.5 M{\odot}$, but it has little effect above this value .
 } \label{MF_bin05}
 \end{center}
\end{figure}

We first derive the system mass function that is the mass function for binaries and single stars. It is shown in Fig.~\ref{MF_bin05} as $\alpha_{SYS}$. Then, in order to compare we calculate the mass function for the single stars and the components of the binaries ($\alpha_{S}$).

As can be seen in Fig.~\ref{MF_bin05}, the slope of the mass function in the low-mass range decreases as the binary fraction increases, while there is no significant change in the high-mass end which is the observed range in this paper.  A high binary fraction,
say 90\%, which is most probably not realistic for an evolved cluster like Pal 4,
can make the mass
function extremely flattened (i.e. $\alpha_{1SYS}\simeq0.24$) in the low-mass range $m/M{\odot}\leq0.5$.
This confirms the results of \citet{Kroupa92b}, \citet{Kroupa93}, \citet{Kroupa09} and \citet{Khalaj13} that even under extreme circumstances (100\%
binaries or higher order multiples), the effect of unresolved multiple systems on the power-law index of the mass function slope is small ($\leq 0.1$) at the high-mass end.

A decrease of $\alpha$ at the low-mass end can be expected since many low-mass stars will be hidden in binaries with more massive companions. However, the effect at the high-mass end is small, even for a binary fraction of 90\%. With a flattening of $\approx0.2$ it is of the order of the uncertainties in the observations. We therefore conclude that unresolved binaries cannot be responsible for the flattening of the mass function in the observed mass range in Pal~4.

\section{Conclusions}\label{Sec:Conclusions}

This paper is the second study in which we model the dynamical evolution of a Galactic globular cluster over its entire lifetime by direct $N$-body simulations on a star-by star basis. While we focused on Pal~14 in Paper I \citep{Zonoozi11}, we here investigate the diffuse outer halo globular cluster Pal~4 using the $N$-body code \textsc{nbody6} \citep{Aarseth03}.

Recent observational work on Pal 4 \citep{Frank12} has shown that the global mass function slope in the mass range 0.55-0.85 M${\odot}$ is $\alpha=1.4\pm0.25$, i.e.~significantly shallower than a canonical mass function slope of about 2.3 \citep{Kroupa01}. Similar results have been found for a number of Milky Way globular clusters (see, e.g.,  \citealt{De Marchi07,Jordi09,Paust10,Frank12,Hamren13}).
Interestingly, \cite{Frank12} also found that the slope of the mass function
steepens with radius from a slope of $\alpha \leq 1$ inside about $1.3 r_h$ to $\alpha \geq 2.3$ at the largest observed radii, indicating the presence of mass segregation in Pal~4 and therefore constraining numerical models much more than our previous target Pal~14 could do.

A preferential loss of low-mass stars due to two-body relaxation would be a natural explanation for the observed mass function depletion \citep{Baumgardt03}.
However, for diffuse outer halo clusters such as Pal~4 and Pal~14  (i.e., a low mass together with a large half-mass radius),  the present-day two-body relaxation time is of the order of a Hubble time. Therefore, relaxation should be inefficient in these clusters and the observations should be an indication for primordial mass segregation. Alternatively the cluster could have been more compact in the past such that relaxation was more important at that time. To test these scenarios, we have tried to find the best possible evolutionary model for Pal~4 by running a set of models with varying initial
half-mass radii and total masses, until we got an adequate fit to the observed structural parameters.

While it is relatively straightforward to find initial models which reproduce the observed structural parameters of Pal~4, i.e. half-light radius, total mass and velocity dispersion, it is very difficult if not impossible to reproduce its global mass function and degree of mass segregation. Because the models have to start with a comparatively low mass and large half-mass radius of about 55000 M${\odot}$ and 10 pc, low-mass star depletion and mass segregation are very ineffective in these clusters.
We showed that models evolving on circular orbits, starting with a Kroupa IMF, and without primordial mass segregation do not produce enough depletion in the slope of the mass function. In addition, these models do not develop enough mass segregation within the cluster lifetime to match the observations.
It should be noted that the current conclusions are based on
the assumption of a circular orbit for Pal~4. The orbit of Pal 4 is unknown however. In the case of an eccentric orbit the Galactic field changes with time, which could significantly affect the dynamical evolution of Pal 4.

We also find that the present-day global mass function slope of Pal 4 cannot be reproduced in models starting with a canonical but primordially segregated IMF, not even by using very high degrees of primordial segregation.
Models starting with a flattened IMF reach enough depletion in the global mass function to be compatible with the observations. However, the radial variation of the mass function slope is significantly better reproduced when we include both a flattened IMF \textit{and} primordial mass segregation.

This is similar to our findings from Paper I \citep{Zonoozi11}, where we
concluded that Pal~14 must have undergone one of two scenarios:
\begin{enumerate}
\item the observed mass function may be a result of dynamical evolution starting from a canonical Kroupa IMF with a high degree of primordial mass segregation;\\
\item the observed mass function may be the result of an already established non-canonical IMF depleted in low-mass stars, which might have been obtained during a violent early phase of gas-expulsion of an embedded cluster with primordial mass segregation \citep{Marks10}.
\end{enumerate}
Now, for Pal~4 we can exclude the first scenario as we have observations covering larger parts of the cluster and hence have more precise knowledge of the present-day global mass function and the degree of mass segregation.

This leaves us with the assumption that the peculiar mass function and the cluster's unusual extent have been imprinted on Pal~4 during its very early lifetime. The inferred initial half-light radius of about 10 pc is significantly larger than the present-day half-light radii of most globular clusters, which are narrowly distributed around 3 pc \citep{Jordan05}. This could be a footprint of the weaker external tidal force of Pal~4's host galaxy during its formation. The Galactic tidal field, which we here model as being static, has evolved significantly since Pal~4's birth and might have been much weaker 11 Gyr ago. The cluster might have also been born in a now detached/disrupted dwarf galaxy.

Alternatively, since star clusters lose more mass during pericentric passages on eccentric orbits, and undergo stronger expansion due to the weaker tidal fields at larger Galactic radii \citep{Madrid12}, an eccentric cluster orbit might have had an important influence on Pal~4's evolution, as it could have had a much smaller initial size and significantly higher mass. We leave this scenario for an upcoming paper to be investigated in detail (K\"upper et al., in preparation).

\section*{Acknowledgements}
HB acknowledges support from the Australian Research Council through Future Fellowship grant FT0991052.
AHWK would like to acknowledge support through the DFG Research Fellowship KU 3109/1-1, and support from NASA through Hubble Fellowship grant HST-HF-51323.01-A awarded by the Space Telescope Science Institute, which is operated by the Association of Universities for Research in Astronomy, Inc., for NASA, under contract NAS 5-26555. MJF gratefully acknowledges support from the DFG via Emmy Noether Grant Ko 4161/1.

\bsp \label{lastpage} 
\begin{thebibliography}{99}

\bibitem[\protect\citeauthoryear{Aarseth}{2003}]{Aarseth03}
Aarseth S.~J., 2003, Gravitational N-Body Simulations, Cambridge University Press, Cambridge

\bibitem[\protect\citeauthoryear{Allen \& Santillan}{1991}]{Allen91}
Allen C., Santillan A., 1991, Rev. Mex. Astron. Astrofis, 22, 255

\bibitem[\protect\citeauthoryear{Allison et al.}{2009}]{Allison09} Allison R.~J., Goodwin S.~P., Parker
R.~J., de Grijs R., Portegies Zwart S.~F., Kouwenhoven M.~B.~N., 2009, ApJ, 700, L99

\bibitem[\protect\citeauthoryear{Banerjee
\& Kroupa}{2013}]{Banerjee13} Banerjee S., Kroupa P., 2013, ApJ, 764, 29

\bibitem[\protect\citeauthoryear{Bastian et al.}{2008}]{Bastian08} Bastian N., Gieles M., Goodwin S.~P.,
Trancho G., Smith L.~J., Konstantopoulos I., Efremov Y., 2008, MNRAS, 389, 223

\bibitem[\protect\citeauthoryear{Baumgardt \& Makino}{2003}]{Baumgardt03}
Baumgardt H., Makino J., 2003, MNRAS, 340, 227.


\bibitem[\protect\citeauthoryear{Baumgardt \& Kroupa}{2007}]{Baumgardt07}
Baumgardt H., Kroupa P., 2007, MNRAS, 380, 1589

\bibitem[\protect\citeauthoryear{Baumgardt, De Marchi, \& Kroupa}{2008}]{Baumgardt08a}
Baumgardt H., De Marchi G., Kroupa P., 2008, ApJ, 685, 247

\bibitem[\protect\citeauthoryear{Baumgardt, Kroupa, \& Parmentier}{2008}]{Baumgardt08b}
Baumgardt H., Kroupa P., Parmentier G., 2008, MNRAS, 384, 1231


\bibitem[\protect\citeauthoryear{Bonnell et al.}{1997}]{Bonnell97} Bonnell I.~A., Bate M.~R., Clarke C.~J.,
Pringle J.~E., 1997, MNRAS, 285, 201

\bibitem[\protect\citeauthoryear{Bonnell et al.}{2001}]{Bonnell01} Bonnell I.~A., Clarke C.~J., Bate M.~R.,
Pringle J.~E., 2001, MNRAS, 324, 573

\bibitem[\protect\citeauthoryear{Bonnell \& Davies}{1998}]{Bonnell98}
Bonnell I. A. \& Davies M. B., 1998, MNRAS, 295, 691

\bibitem[\protect\citeauthoryear{Bonnell \& Bate}{2006}]{Bonnell06} Bonnell I.~A., Bate M.~R., 2006, MNRAS, 370, 488

\bibitem[\protect\citeauthoryear{Burbidge \& Sandage}{1958}]{Burbidge58} Burbidge E.~M., Sandage A., 1958, ApJ, 127, 527

\bibitem[\protect\citeauthoryear{Christian \& Heasley}{1986}]{Christian86} Christian C.~A., Heasley J.~N., 1986, ApJ, 303, 216

\bibitem[\protect\citeauthoryear{de Grijs et al.}{2002}]{de Grijs02} de Grijs R., Gilmore G.~F., Johnson R.~A.,
Mackey A.~D., 2002, MNRAS, 331, 245

\bibitem[\protect\citeauthoryear{de Grijs}{2010}]{de Grijs10} de Grijs R., 2010, Philos. Trans. R. Soc. A: Math. Phys. Eng. Sci., 368, 693

\bibitem[\protect\citeauthoryear{Dabringhausen, Fellhauer, \& Kroupa}{2010}]{Dabringhausen10}
Dabringhausen, J., Fellhauer, M., \& Kroupa, P. 2010, MNRAS, 403, 1054

\bibitem[\protect\citeauthoryear{De Marchi et al.}{2007}]{De Marchi07} De Marchi G., Paresce F., Pulone L., 2007, ApJ, 656, L65

\bibitem[\protect\citeauthoryear{Dotter et al.}{2008}]{Dotter08} Dotter A., Chaboyer B., Jevremovi{\'c} D., Kostov V., Baron E., Ferguson J.~W., 2008, ApJS, 178, 89

\bibitem[\protect\citeauthoryear{Fellhauer, Wilkinson, \& Kroupa}{2009}]{Fellhauer09} Fellhauer M., Wilkinson M.~I., Kroupa P., 2009, MNRAS, 397, 954

\bibitem[\protect\citeauthoryear{Fischer et al.}{1998}]{Fischer98} Fischer P., Pryor C., Murray S., Mateo M.,
Richtler T., 1998, AJ, 115, 592

\bibitem[\protect\citeauthoryear{Frank et al.}{2012}]{Frank12} Frank M.~J., Hilker M., Baumgardt H., C{\^o}t{\'e} P., Grebel
E.~K., Haghi H., K{\"u}pper A.~H.~W., Djorgovski S.~G., 2012, MNRAS, 423, 2917

\bibitem[\protect\citeauthoryear{Giersz \& Heggie}{2011}]{Giersz11}
Giersz M., Heggie D.~C., 2011, MNRAS, 410, 2698

\bibitem[\protect\citeauthoryear{Gouliermis et al.}{2004}]{Gouliermis04} Gouliermis D., Keller S.~C., Kontizas M., Kontizas E., Bellas-Velidis I., 2004, A\&A, 416, 137

\bibitem[\protect\citeauthoryear{Gouliermis, de Grijs, \& Xin}{2009}]{Gouliermis09} Gouliermis D.~A., de Grijs R., Xin Y., 2009, ApJ, 692, 1678

\bibitem[\protect\citeauthoryear{G{\"u}rkan, Freitag, \& Rasio}{2004}]{Gurkan04} G{\"u}rkan M.~A., Freitag M., Rasio F.~A., 2004, ApJ, 604, 632


\bibitem[\protect\citeauthoryear{Hamren et al.}{2013}]{Hamren13}
Hamren K.~M., Smith G.~H., Guhathakurta P., Dolphin A.~E., Weisz D.~R.,
Rajan A., Grillmair C.~J., 2013, AJ, 146, 116

\bibitem[\protect\citeauthoryear{Harris}{1996}]{Harris96}
Harris W. E., 1996, AJ, 112, 1487 (arXiv:1012.3224)

\bibitem[\protect\citeauthoryear{Heggie \& Hut}{2003}]{Heggie03}
Heggie D., Hut P., 2003, The Gravitational Million Body Problem. Cambridge University Press, Cambridge

\bibitem[\protect\citeauthoryear{Hillenbrand}{1997}]{Hillenbrand97}
Hillenbrand L.~A., 1997, AJ, 113, 1733

\bibitem[\protect\citeauthoryear{Hillenbrand \& Hartmann}{1998}]{Hillenbrand98} Hillenbrand L.~A., Hartmann L.~W., 1998, ApJ, 492, 540

\bibitem[\protect\citeauthoryear{Hurley, Pols \& Tout}{2000}]{Hurley00}
Hurley J.~R., Pols O.~R., Tout C.~A., 2000, MNRAS, 315, 543

\bibitem[\protect\citeauthoryear{Hurley, Tout \& Pols}{2002}]{Hurley02}
Hurley J.~R., Tout C.~A., Pols O.~R., 2002, MNRAS, 329, 897

\bibitem[\protect\citeauthoryear{Irrgang et al.}{2013}]{Irrgang13}
Irrgang A.,  Wilcox B., Tucker E., Schiefelbein L., 2013, A\&A, 549, A137

\bibitem[\protect\citeauthoryear{Jord{\'a}n et al.}{2005}]{Jordan05}
Jord{\'a}n A., et al., 2005, ApJ, 634, 1002

\bibitem[\protect\citeauthoryear{Jordi et al.}{2009}]{Jordi09}
Jordi K. et al., 2009, AJ, 137, 4586

\bibitem[\protect\citeauthoryear{Khalaj \& Baumgardt}{2013}]{Khalaj13}
Khalaj P., Baumgardt B., 2013, MNRAS, 434, 3236

\bibitem[\protect\citeauthoryear{King}{1966}]{King66} King I.~R., 1966, AJ, 71, 276

\bibitem[\protect\citeauthoryear{Klessen}{2001}]{Klessen01}
Klessen R.~S., 2001, ApJ, 556, 837

\bibitem[\protect\citeauthoryear{Kroupa, Gilmore, \& Tout}{1991}]{Kroupa91} Kroupa P., Gilmore G., Tout C.~A., 1991, MNRAS, 251, 293

\bibitem[\protect\citeauthoryear{Kroupa \& Tout}{1992}]{Kroupa92}
Kroupa P., \& Tout C.~A., 1992, MNRAS, 259, 223

\bibitem[\protect\citeauthoryear{Kroupa, Gilmore,
\& Tout}{1992}]{Kroupa92b} Kroupa P., Gilmore G., Tout C.~A., 1992, AJ, 103, 1602

\bibitem[\protect\citeauthoryear{Kroupa, Tout, \& Gilmore}{1993}]{Kroupa93} Kroupa P., Tout C.~A., Gilmore G., 1993, MNRAS, 262, 545

\bibitem[\protect\citeauthoryear{Kroupa}{1995a}]{Kroupa95}
Kroupa P., 1995a, MNRAS, 277, 1491
%
\bibitem[\protect\citeauthoryear{Kroupa}{1995b}]{Kroupa95b}
Kroupa P., 1995b, MNRAS, 277, 1507

\bibitem[\protect\citeauthoryear{Kroupa}{1995c}]{Kroupa95c}
Kroupa P., 1995c, MNRAS, 277, 1522

\bibitem[\protect\citeauthoryear{Kroupa}{2001}]{Kroupa01}
Kroupa P., 2001, MNRAS, 322, 231

\bibitem[\protect\citeauthoryear{Kroupa, Aarseth,
\& Hurley}{2001}]{Kroupa01b} Kroupa P., Aarseth S., Hurley J., 2001, MNRAS, 321, 699

\bibitem[\protect\citeauthoryear{Kroupa}{2002}]{Kroupa02}
Kroupa P., 2002, Science, 295, 82

\bibitem[\protect\citeauthoryear{Kroupa}{2008}]{Kroupa08}
Kroupa, P. 2008, in  Aarseth S. J.,  Tout C. A.,  Mardlingin R. A., eds, Lecture Notes in Physics,  Vol. 760, The Cambridge N-Body Lectures. Springer-Verlag, Berlin, p. 181

\bibitem[\protect\citeauthoryear{Kroupa et al.}{2013}]{Kroupa13}
Kroupa P., Weidner ., Pflamm-Altenburg J., Thies I., Dabringhausen J., Marks M., Maschberger T., 2013, Stellar Systems and Galactic Structure, Vol. 5. Springer-Verlag, Berlin

\bibitem[\protect\citeauthoryear{K{\"u}pper et al.}{2011}]{Kupper11}
K{\"u}pper A.~H.~W., Maschberger T., Kroupa P., Baumgardt H., 2011, MNRAS, 417, 2300

\bibitem[\protect\citeauthoryear{McMillan, Vesperini, \& Portegies Zwart}{2007}]{McMillan07} McMillan S.~L.~W., Vesperini E., Portegies Zwart S.~F., 2007, ApJ, 655, L45

\bibitem[\protect\citeauthoryear{Madrid, Hurley, \& Sippel}{2012}]{Madrid12}
Madrid J.~P., Hurley J.~R., Sippel A.~C., 2012, ApJ, 756, 167

\bibitem[\protect\citeauthoryear{Marks, Kroupa, \& Baumgardt}{2008}]{Marks08}
Marks M., Kroupa P., Baumgardt H., 2008, MNRAS, 386, 2047

\bibitem[\protect\citeauthoryear{Marks \& Kroupa}{2010}]{Marks10}
Marks M., Kroupa P., 2010, MNRAS, 406, 2000

\bibitem[\protect\citeauthoryear{Moeckel \& Bonnell}{2009}]{Moeckel09} Moeckel N., Bonnell I.~A., 2009, MNRAS, 400, 657

\bibitem[\protect\citeauthoryear{Mouri \& Taniguchi}{2002}]{Mouri02} Mouri H., Taniguchi Y., 2002, ApJ, 566, L17

\bibitem[\protect\citeauthoryear{Nitadori \& Aarseth}{2012}]{Nitadori12}
Nitadori K., Aarseth S.~J., 2012, MNRAS, 424, 545

\bibitem[\protect\citeauthoryear{Paust et al.}{2010}]{Paust10} Paust N.~E.~Q., et al., 2010, AJ, 139, 476

\bibitem[\protect\citeauthoryear{Parmentier et al.}{2008}]{Parmentier08}
Parmentier, G., Goodwin, S. P., Kroupa, P., Baumgardt, H., 2008,
ApJ, 678, 347


\bibitem[\protect\citeauthoryear{Plummer}{1911}]{Plummer11}
Plummer H.~C., 1911, MNRAS, 71, 460

\bibitem[\protect\citeauthoryear{Sabbi et al.}{2008}]{Sabbi08}
Sabbi E., et al., 2008, AJ, 135, 173

\bibitem[\protect\citeauthoryear{Sakamoto et al.}{2003}]{Sakamoto03}
Sakamoto, T., Chiba, M., \& Beers, T. C. 2003, A\&A, 397, 899

\bibitem[\protect\citeauthoryear{Salpeter}{1955}]{Salpeter55}
Salpeter E.~E., 1955, ApJ, 121, 161

\bibitem[\protect\citeauthoryear{Sana \& Evans}{2011}]{Sana11}
Sana H., Evans C.~J., 2011, in Neiner C., Wade G., Meynet G., Peters G.,
eds, Proc. IAU Symp. 272, Active OB Stars: Structure, Evolution, Mass
Loss, and Critical Limits. Cambridge Univ. Press, Cambridge, p. 474

\bibitem[\protect\citeauthoryear{Sirianni et
al.}{2002}]{Sirianni02} Sirianni M., Nota A., De Marchi G.,
Leitherer C., Clampin M., 2002, ApJ, 579, 275

\bibitem[\protect\citeauthoryear{Stolte et al.}{2006}]{Stolte06}
Stolte A., Brandner W., Brandl B., Zinnecker H., 2006, AJ, 132, 253

\bibitem[\protect\citeauthoryear{VandenBerg}{2000}]{VandenBerg00}
VandenBerg D.~A., 2000, ApJS, 129, 315

\bibitem[\protect\citeauthoryear{Vesperini \& Heggie}{1997}]{Vesperini97}
Vesperini E. \& Heggie D. C., 1997, MNRAS, 289, 898

\bibitem[\protect\citeauthoryear{Vesperini et al.}{2009}]{Vesperini09b}
Vesperini E., McMillan S., Portegies Zwart S., 2009b, ApJ, 698, 615

\bibitem[\protect\citeauthoryear{Vogt et al.}{1994}]{Vogt94}
Vogt S. S., et al., 1994, Proc. SPIE, 2198, 362

\bibitem[\protect\citeauthoryear{Weidner, Kroupa,
\& Maschberger}{2009}]{Kroupa09} Weidner C., Kroupa P., Maschberger T., 2009, MNRAS, 393, 663

\bibitem[\protect\citeauthoryear{Wilkinson \& Evans}{1999}]{Wilkinson99}
Wilkinson, M. I. \& Evans, N. W. 1999, MNRAS, 310, 645

\bibitem[\protect\citeauthoryear{Wilkinson et al.}{2003}]{Wilkinson03} Wilkinson M.~I., Hurley J.~R., Mackey
A.~D., Gilmore G.~F., Tout C.~A., 2003, MNRAS, 343, 1025

\bibitem[\protect\citeauthoryear{Zonoozi et al.}{2011}]{Zonoozi11}
Zonoozi A.~H., K{\"u}pper A.~H.~W., Baumgardt H., Haghi H., Kroupa P., Hilker M., 2011, MNRAS, 411, 1989 



\end{thebibliography}
\end{document}